\documentclass[final,onefignum,onetabnum]{siamart171218}

\usepackage{lipsum}
\usepackage{amsfonts}
\usepackage{graphicx}
\usepackage{epstopdf}
\usepackage{algorithmic}
\ifpdf
  \DeclareGraphicsExtensions{.eps,.pdf,.png,.jpg}
\else
  \DeclareGraphicsExtensions{.eps}
\fi

\usepackage{nicefrac}
\usepackage{bm}
\usepackage{bbm}
\usepackage{tikz}
\usetikzlibrary{decorations.pathmorphing}
\usetikzlibrary{matrix}

\usepackage{mathtools}
\usepackage{multirow, makecell}
\usepackage{subcaption}

\usepackage{enumitem}
\setlist[enumerate]{leftmargin=.5in}
\setlist[itemize]{leftmargin=.5in}

\newsiamremark{remark}{Remark}
\newsiamremark{hypothesis}{Hypothesis}
\crefname{hypothesis}{Hypothesis}{Hypotheses}
\newsiamthm{claim}{Claim}

\headers{Dynamic Ecological System Measures}{Huseyin Coskun}

\title{Dynamic Ecological System Measures} 
\subtitle{A Holistic Analysis of Compartmental Systems}

\author{Huseyin Coskun\thanks{Department of Mathematics, University of Georgia, Athens, GA 30602 (\email{hcoskun@uga.edu}).} }

\usepackage{amsopn}
\DeclareMathOperator{\diag}{diag}
\DeclareMathOperator{\sgn}{sgn}

\ifpdf
\hypersetup{
  pdftitle={Dynamic Ecological System Measures},
  pdfauthor={Huseyin Coskun}
}
\fi

\begin{document}

\maketitle

\begin{abstract}
The system decomposition theory has recently been developed for the dynamic analysis of nonlinear compartmental systems. The application of this theory to the ecosystem analysis has also been introduced in a separate article. Based on this methodology, multiple new dynamic ecological system measures and indices of matrix, vector, and scalar types are systematically introduced in the present paper. These mathematical system analysis tools are quantitative ecological indicators that monitor the flow distribution and storage organization, quantify the direct, indirect, acyclic, cycling, and transfer (\texttt{diact}) effects and utilities of one compartment on another, identify the system efficiencies and stress, measure the compartmental exposures to system flows, determine the residence times and compartmental activity levels, and ascertain the system resilience and resistance in the case of disturbances. The proposed dynamic system measures and indices, thus, extract detailed information about ecosystems' characteristics, as well as their functions, properties, behaviors, and various other system attributes that are potentially hidden in and even obscured by data. A dynamic technique for the quantitative characterization and classification of main interspecific interactions and the determination of their strength within food webs is also developed based on the \texttt{diact} effect and utility indices. Moreover, major concepts and quantities in the current static network analyses are also extended to nonlinear dynamic settings and integrated with the proposed dynamic measures and indices in this unifying mathematical framework. Therefore, the proposed methodology enables a holistic view and analysis of ecological systems. We consider that the proposed methodology brings a novel complex system theory to the service of urgent and challenging environmental problems of the day and has the potential to lead the way to a more formalistic ecological science.
\end{abstract}

\begin{keywords}
system decomposition theory, complex systems theory, dynamic ecological network analysis, nonlinear dynamic compartmental systems, dynamic system and subsystem partitioning, \texttt{diact} flows and storages, \texttt{diact} effect measures and indices, \texttt{diact} system efficiencies and stress, \texttt{diact} utility measures and indices, \texttt{diact} exposures and residence times, system resilience, system resistance, dynamic cycling index, dynamic indirect effects, food webs, interspecific interactions, dynamic input-output analysis, dynamic input-output economics, socio-economic systems, epidemiology, infectious diseases, toxicology, pharmacokinetics, neural networks, chemical and biological systems, control theory, information theory, information diffusion, social networks, computer networks, malware propagation, graph theory, traffic flow
\end{keywords}

\begin{AMS}
34A34, 35A24, 37C60, 37N25, 37N40, 70G60, 91B74, 92B20, 92C42, 92D30, 92D40, 93C15, 94A15
\end{AMS}

\section{Introduction}
\label{sec:intro}

Environmental problems have been a common topic of scholarly conversation for decades. As environmental issues persist and proliferate, the language and methods through which these problems are examined also evolve. Although traditional ecology has been used effectively in dealing with a variety of complex environmental problems, the field remains largely descriptive in nature. It has yet to arrive at a formal theory and methodology for analyzing the complex relationships between organisms and their environment or man and nature.

Ecosystems are natural systems made up of living and non-living components. Ecosystem ecology deals with interactions between species and their physical environment. More specifically, this interdisciplinary science studies the flows of energy and matter between the biotic and abiotic components of ecosystems based on conservation principles. Ecosystem ecology plays an important role in understanding current global environmental problems and determining how local mechanisms interact with these problems. Enhancements in dealing with environmental issues will ultimately depend on advances in such basic sciences. Mathematical theories and modeling are at the forefront of continued endeavors leading to a more formalistic and theoretical ecological science devoted to the discovery of basic scientific laws. Compartmental models are generally used for mathematical abstractions of ecological systems where the compartments represent ecosystem components. Within this mathematical framework, system measures and indices are then formulated to serve as quantitative ecological indicators.

Ecological models are widely analyzed in the literature, but the current methodologies are developed for special cases, such as linear systems and static models. Ecological networks and complexity in living systems are analyzed, for example, at steady state in the context of {\em information theory} and {\em thermodynamics} \cite{Ulanowicz1972,Hirata1985,Ulanowicz2004,Ulanowicz2013}, as well as the {\em hierarchy theory} \cite{Allen2014}. Building on economic {\em input-output analysis} of \cite{Leontief1936, Leontief1966}, introduced into ecology by \cite{Hannon1973}, another static approach called the {\em environ theory} has been developed based on conservation principles \cite{Patten1978,Matis1981}. Several software developments computerize these static methods \cite{Ulanowicz1991,Christensen1992, Fath2006, Kazanci2009a, Schramski2011, Borrett2014c}.

Despite the fact that major environmental problems of the day involve change, the dynamic analysis of nonlinear compartmental systems has remained a long-standing, open problem. Current ecosystem measures and indices are all formulated for static systems. For example, Finn's cycling index\textemdash a celebrated ecosystem measure that quantifies cycling system flows\textemdash defined in static ecological network analyses over four decades ago, has still not been made applicable to ecosystem models that change over time \cite{Finn1976}. The indirect effects in ecosystems have also long been a well-established empirical fact \cite{Paine1966,Strauss1991,Wootton1993,Menge1995,Menge1997,Wootton2002}. Theoretical ecological explorations of the concept began as early as the 1970s \cite{Holt1977,Patten1992,Fath1999,Ma2013}. The indirect effects is an important concept for also many other fields, such as network theory, graph theory, and neural networks, but it is one of the main topics in the input-output economics~\cite{Leontief1966}. Although the indirect effects have been a topic of scholarly conversation for about a century, they have never been formulated for dynamic systems before. Therefore, there is an urgent need for dynamic methods and measures for nonlinear ecological system analysis \cite{Chen2001,Hastings1996,Hastings2004}.

The indirect effects are particularly important for the characterization and classification of interspecific interactions within food webs. The classification through direct relationships alone can turn out to be incorrect without holistically considering the entire network of interactions. Moreover, the conditions and states of communities in food webs, such as extinction, can be dynamically regulated by the temporal variations and seasonal shifts \cite{Dell2006,Zanden2016}. Community ecology qualitatively describes interspecific interactions using network topology. On the other hand, for complex networks, such characterization becomes increasingly difficult, if possible at all \cite{Wootton1994,Menge1995,Menge1997,Holt1997,Patten2007}. Multiple food chains of potentially different lengths between two species, for example, disallow the graph-theoretical classification based on the length of the chains between two species. Some parametric characterizations are proposed in the literature \cite{Ulanowicz1990b,Patten1992,Fath1998,Fath2007,Tuominen2014}, however, they are only for static systems and have some disadvantages due to the method formulations as detailed by \cite{Coskun2017SESM,Coskun2019ITR}. 

The \textit{system decomposition theory} and comprehensive methods recently developed by \cite{Coskun2017DCSAM,Coskun2017NDP} for the dynamic analysis of nonlinear compartmental systems potentially addresses the disconnect between current static and computational methods and applied ecological needs. The system decomposition theory is based on the novel analytical and explicit, mutually exclusive and exhaustive \textit{system} and \textit{subsystem partitioning methodologies}. While the dynamic {\em system} partitioning provides the \emph{subthroughflows} and \emph{substorages} to determine the distribution of the initial stocks and environmental inputs, as well as the organization of the associated storages derived from these stocks and inputs individually and separately within the system, the {\em subsystem} partitioning yields the {\em transient flows} and \emph{storages} to determine the distribution of arbitrary intercompartmental flows and the organization of associated storages along any given flow path within the subsystems. Indirect transactions between any two system compartments have never been formulated before, even for static ssytems. Graph theoretically, a nonzero indirect flow between two compartments indicates the existence of an indirect path between these compartments. Not only the indirect flows, but the dynamic {\em direct}, {\em indirect}, {\em acyclic}, {\em cycling}, and {\em transfer} ($\texttt{diact}$) flows and associated storages transmitted from one compartment, directly or indirectly, to any other are also analytically characterized, systematically classified, and mathematically formulated for the first time. Consequently, through the system decomposition theory, the evolution of the initial stocks, environmental inputs, and arbitrary intercompartmental system flows, as well as the associated storages derived from these stocks, inputs, and flows can be tracked individually and separately within the system.

The system decomposition theory constructs a foundation for the development of new mathematical system analysis tools as quantitative ecosystem indicators. Based on this theory, multiple measures and indices of matrix, vector, and scalar types for the dynamic $\texttt{diact}$ effects, utilities, exposures, and residence times, as well as the corresponding system efficiencies, stress, resilience and resistance are introduced systematically in the present paper for the first time in literature. The proposed dynamic system measures and indices monitor the flow distribution and storage organization, quantify the \texttt{diact} effects and utilities of one compartment directly or indirectly on another, identify the system efficiencies and stress, measure the compartmental exposures to system flows, determine the residence times and compartmental activity levels, and ascertain the system resilience and resistance in the case of disturbances. As a result, these measures and indices dynamically quantify ecosystems' characteristics, including their functions, features, properties, and various other system attributes that are potentially hidden in and even obscured by data. They ultimately enable the characterization and classification of ecosystems, precise analyses of system structure and behavior, as well as a detailed understanding of the dynamics of individual system compartments. Therefore, they may prove useful also for environmental assessment and management. 

A novel mathematical technique for the quantitative characterization and classification of the main interspecific interaction types, and notably, for the determination of their strength is also developed in the present manuscript. This technique that uses system flows and storages for the quantitative characterization of interspecific interactions sets up a bridge between two main branches of ecology: ecosystem ecology and community ecology.  Consequently, the proposed methodology, as a whole, leads to a holistic analysis of ecosystems and serves as a quantitative platform for testing empirical hypotheses, ecological inferences, and, potentially, theoretical developments. In effect, the proposed methodology brings a novel, formal, deterministic, complex system theory to the service of urgent and challenging environmental problems of the day and has the potential to lead the way to a more formalistic ecological science.

The system decomposition theory and its holistic nature has recently been elaborated further for static systems by \cite{Coskun2017SCSA,Coskun2017SESM}. In these studies, the current static measures and indices were reformulated with a different derivation rationale in the context of the system decomposition and, therefore, were integrated with the novel system analysis tools developed through the proposed methodology. Unique relationships among some current static measures were unveiled and corrections on some existing formulations were suggested. The input- and output-oriented ecosystem analysis were also integrated, and remarkably, their duality was established through some similarity relationships. The system flows and storages have been treated separately in the static ecological network analyses. The system decomposition theory integrates flow and storage partitioning effectively through the novel concept of residence time. Consequently, various components of ecosystem analyses are effectively combined and holistically integrated within the proposed unifying mathematical framework. The current compartmental level system analyses were also refined to subcompartmental level to address the full complexity of both dynamic and static ecological systems. The system decomposition theory, thus, refines system analyses from the current static, linear, compartmental level to the dynamic, nonlinear, subcompartmental level.

The system decomposition theory is applicable to any compartmental system regardless of its nature, whether naturogenic or anthropogenic. It can be used, for example, to analyze mass or energy transfers between species of different trophic levels in a complex network or along a given food chain of a food web \cite{Belgrano2005}. The proposed methodology can also be used for the analysis of models designed for material flows in industry or the dynamics of the terrestrial carbon cycle on the regional and global scale \cite{Bailey2004, Rasmussen2016}. Although the motivating applications for this paper are ecological and environmental, the applicability of the proposed methods extend to other realms such as economics, pharmacology, epidemiology, chemical reaction kinetics, biomedical systems, neural networks, social networks, and information science\textemdash in fact, wherever dynamic compartmental models of conserved quantities can be constructed. In the context of economics, in particular, the system decomposition theory can be considered as the mathematical foundation of the \textit{dynamic input-output economics} \cite{Coskun2017NDP}.

The proposed methodology is applied to several ecosystem models in Section~\ref{sec:results} to illustrate the efficiency and wide applicability of the proposed measures and indices. These models have recently been analyzed for their flow and storage distributions and intercompartmental transfer dynamics through system and subsystem partitioning methodologies~\cite{Coskun2017DCSAM,Coskun2017SCSA}. In the present manuscript, the measures and indices for the dynamic \texttt{diact} effects, utilities, exposures, and residence times, as well as the system efficiencies, stress, resilience, and resistance are presented for these ecosystems. The interspecific interactions in some models together with their strength are also analyzed though the proposed mathematical classification technique.

The paper is organized as follows: the mathematical formulations of the ecological system measures and indices are introduced in Section~\ref{sec:method}, results and examples are provided in Section~\ref{sec:results}, and discussion and conclusions follow in Section~\ref{sec:disc} and~\ref{sec:conc}.

\section{Methods}
\label{sec:method}

The \textit{system decomposition theory} has recently been developed for the dynamic analysis of nonlinear compartmental systems by \cite{Coskun2017DCSAM,Coskun2017NDP}. The theory is based on the novel analytical and explicit, mutually exclusive and exhaustive {\em system} and {\em subsystem partitioning methodologies}. While the proposed dynamic {\em system} partitioning yields the subthroughflow and substorage matrices  to determine the distribution of the initial stocks and environmental inputs, as well as the organization of the associated storages generated by these stock and inputs individually and separately within the system, the {\em subsystem} partitioning yields the transient flows and storages along a particular flow path to determine the distribution of arbitrary intercompartmental flows and the organization of the associated storages generated by these flows within the subsystems. Consequently, the evolution of the initial stocks, environmental inputs, and arbitrary intercompartmental system flows, as well as the associated storages derived from these stocks, inputs, and flows can be tracked individually and separately within the system. The system decomposition, therefore, yields the decomposition of system flows and storages to the utmost level, as summarized further below.

Based on the system decomposition, the dynamic direct, indirect, acyclic, cycling, and transfer (\texttt{diact}) flows and associated storages transmitted from one compartment, directly or indirectly, to any other  within the system are also analytically characterized, systematically classified, and mathematically formulated to ascertain the dynamics of intercompartmental transactions. 

The proposed methodology constructs a base for the development of mathematical system analysis tools as quantitative ecosystem indicators. Multiple such novel measures and indices of matrix, vector, and scalar types for the dynamic $\texttt{diact}$ effects, utilities, exposures, and residence times, as well as the corresponding system efficiencies, stress, resilience, and resistance are introduced systematically in this section. The static versions of these measures and indices are developed in a separate article~\cite{Coskun2017SCSA}. A mathematical technique for the dynamic analysis of food webs and food chains is also introduced at the end of this section. Unlike the current qualitative approaches, this technique proposes a quantitative characterization and classification procedure for main interspecific interaction types and the determination of their strength based on the \texttt{diact} effects and utilities.

The standard governing equations for compartmental dynamics are
\begin{equation}
\label{eq:model_c1}
\dot{x}_i(t) = \left( z_i(t,x) + \sum_{\substack{j=1}}^n f_{ij}(t,x)  \right) - \left( y_i(t,x) + \sum_{\substack{j=1 }}^n f_{ji}(t,x) \right)
\end{equation} 
with the initial conditions of $x_i(t_0) = x_{i,0}$, for $i=1,\ldots,n$. The terminology and notations used in this equation and throughout the paper are adopted from \cite{Coskun2017DCSAM,Coskun2017NDP}: 
\begin{center} 
    \begin{tabular}[h]{ l p{.7\textwidth} }
\hfill \\    
    $n$ & number of compartments \\
    $t$ & time $[\mathrm{t}]$ \\
    $x_i(t)$ & total material (mass) $[\mathrm{m}]$  (or energy, currency) in compartment $i$, $i = 1,\ldots,n$, at time $t$ \\
    $f_{ij}(t,x)$ & nonnegative flow from compartment $j$ to $i$, at time $t$ $[\mathrm{m}/\mathrm{t}]$ \\    
    $y_{i}(t,x) = f_{0{i}}(t,x)$ & environmental ($j=0$) output from compartment $i$ at time $t$ \\ 
     $z_{i}(t,x) = f_{{i}0}(t,x)$ & environmental input into compartment $i$ at time $t$ \\
     \hfill
    \end{tabular}
\end{center} 

For notational convenience, we define a {\em direct flow matrix} function $F$ of size $n \times n$, whose $(i,j)-$element is the nonlinear flow rate from compartment $j$ to $i$, $f_{ij}(t,x)$, as
\begin{equation}
\begin{aligned}
\label{eq:Fmatrix_1}
F(t,x) = \left(f_{ij}(t,x) \right)
\end{aligned}
\nonumber
\end{equation}
and the {\em total inflow} and {\em outflow vector} functions as
\begin{equation}
\begin{aligned}
\label{eq:Tvector_1}
\check{\tau}(t,x) &= \left[ \check{\tau}_1(t,x), \ldots,\check{\tau}_n(t,x) \right]^T = z(t,x) + F(t,x) \, \mathbf{1}  \quad  \mbox{and} \\ 
\hat{\tau}(t,x) &= \left[ \hat{\tau}_1(t,x),\ldots,\hat{\tau}_n(t,x) \right]^T = y(t,x) + F^T(t,x) \, \mathbf{1} , 
\end{aligned}
\end{equation}
respectively, where 
\begin{equation}
\begin{aligned}
\label{eq:xyzvectors}
x(t) & = [x_1(t),\ldots,x_n(t)]^T , \\ 
z(t,x) &= [z_1(t,x),\ldots,z_n(t,x)]^T \quad \mbox{and} \quad y(t,x)=[y_1(t,x),\ldots,y_n(t,x)]^T 
\end{aligned}
\end{equation}
are the differentiable \emph{state}, {\em input} and {\em output vector} functions, ${\mathbf 1}$ denotes the column vector of size $n$ whose entries are all one, and the superscript $T$ represents the matrix transpose.

The total inflow, $\check{\tau}_i(t,x)$, and outflow, $\hat{\tau}_i(t,x)$, will be called the {\em inward} and {\em outward throughflows} at compartment $i$, respectively, and formulated as
\begin{equation}
\label{eq:in_out_flows}
\check{\tau}_i(t,x) = \sum_{\substack{j=0}}^n f_{ij}(t,x) \quad \mbox{and} \quad  \hat{\tau}_i(t,x) = \sum_{\substack{j=0}}^n f_{ji}(t,x)
\end{equation} 
for $i=1,\ldots,n$. The nonlinear differentiable function $f_{ij}(t,x) \geq 0$ represents nonnegative flow rate from compartment $j$ to $i$ at time $t$. In general, it is assumed that $f_{ii}(t,x) = 0$, but the following analysis is also valid for nonnegative flow from a compartment into itself. Index $j=0$ stands for the environment. We further assume that $f_{ij}(t,x)$ has the following special form:
\begin{equation}
\label{eq:sf}
f_{ij}(t,x) = q^x_{ij}(t,x) \, {x}_j (t)
\end{equation}
where $q^x_{ij}(t,x)$ is a nonlinear function of ${x}$ and $t$, and has the same properties as $f_{ij}(t,x)$ \cite{Coskun2017DCSAM}.

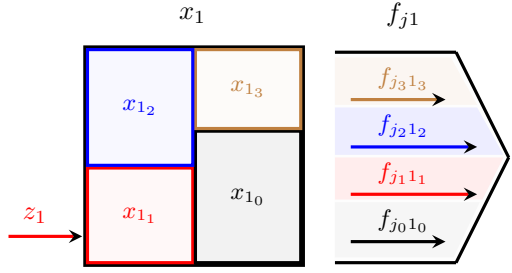
\begin{figure}[t]
\begin{center}
\begin{tikzpicture}[scale=.7]
\centering
    \draw[fill=gray!2]    (4.7,0) -- ++(2.3,0) -- ++(1,2) -- ++(-1,2) -- ++(-2.3,0);           
    \draw[draw=none,fill=red!7]  (4.7,1.2) -- ++ (2.8,0) -- ++ (.4,.8) -- ++ (-3.2,0);             
    \draw[draw=none,fill=gray!7]  (4.7,.1) -- ++ (2.25,0) -- ++ (.5,1.05) -- ++ (-2.75,0);                 
    \draw[draw=none,fill=blue!7]  (4.7,2.05) -- ++ (3.2,0) -- ++ (-.5,.9) -- ++ (-2.7,0);                     
    \draw[draw=none,fill=brown!7]  (4.7,3) -- ++ (2.7,0) -- ++ (-.5,.9) -- ++ (-2.2,0);                         
   \draw[very thick, draw=black] (-.05,-.05) rectangle node(R1) [pos=.5] { } (4.1,4.1) ;
   \draw[very thick, fill=red!3, draw=red, text=red] (0,0) rectangle node(R1) [pos=.5] {\small $x_{1_1}$} (2,1.8) ;
   \draw[very thick, fill=gray!10, draw=black, text=black] (2.05,0) rectangle node(R2) [pos=.5] {\small ${x}_{1_0}$} (4.05,2.5) ;
   \draw[very thick, fill=brown!3, draw=brown, text=brown] (2.05,2.55) rectangle node(R3) [pos=.5] {\small $x_{1_3}$} (4.05,4.05) ;
   \draw[very thick, fill=blue!3, draw=blue, text=blue] (0,1.85) rectangle node(R4) [pos=.5] {\small $x_{1_2}$} (2,4.05) ;      
    \draw[very thick]  (4.7,0) -- (7,0) ;
    \draw[very thick]  (7,0) -- (8,2) ;   
    \draw[very thick]  (8,2) -- (7,4) ;        
    \draw[very thick]  (7,4) -- (4.7,4) ; 
    \draw[very thick,-stealth,draw=red]  (-1.5,.5) -- (-.1,.5) ;     
    \node (z) [text=red] at (-1,0.9) {$z_1$};                        
    \draw[very thick,-stealth,draw=black]  (5,.4) -- (6.8,.4) ;     
    \node (f0) at (6,.8) {\small ${f}_{j_0 1_0}$};      
    \draw[very thick,-stealth,draw=red]  (5,1.3) -- (7.4,1.3) ;         
    \node (f1) [text=red] at (6,1.7) {\small ${f}_{j_1 1_1}$};
    \draw[very thick,-stealth,draw=blue]  (5,2.2) -- (7.4,2.2) ;     
    \node (f2) [text=blue] at (6,2.6) {\small ${f}_{j_2 1_2}$};   
    \draw[very thick,-stealth,draw=brown]  (5,3.1) -- (6.8,3.1) ;     
    \node (f3) [text=brown] at (6,3.5) {\small ${f}_{j_3 1_3}$};                                                     
    \node (x) at (2,4.7) {${x}_{1}$};      
    \node (x) at (6,4.7) {${f}_{j1}$};               
\end{tikzpicture}
\end{center}
\caption{Schematic representation of the dynamic flow partitioning in a three-compartment model system. The figure illustrates subcompartmentalization of compartment $i=1$ and the corresponding dynamic flow partitioning from this compartment to others, $j$.}
\label{fig:sc}
\end{figure}

The system partitioning methodology yields the governing equations for the dynamics of the mutually exclusive subsystems or individual subcompartments as follows (see Figs.~\ref{fig:sd} and ~\ref{fig:sc}): 
\begin{equation}
\label{eq:model_sc_apx}
\dot{x}_{i_k}(t) = \left( z_{i_k}(t,{\rm x}) + \sum_{\substack{j=1}}^n f_{{i_k}{j_k}}(t,{\rm x})  \right) - \left( y_{i_k}(t,{\rm x}) + \sum_{\substack{j=1}}^n f_{{j_k}{i_k}}(t,{\rm x}) \right)
\end{equation} 
for $i=1,\ldots,n$, $k=0,\ldots,n$, with the initial conditions
\begin{equation}
\label{eq:ic_apx}
x_{i_k} (t_0) = \left \{
\begin{aligned}
x_{i,0}, \quad k=0 \\
0,  \quad k \neq 0
\end{aligned}
\right.
\nonumber
\end{equation}
where
\[ {\rm x}(t) = \left[ x_{1_0}(t), \ldots, x_{n_0}(t), x_{1_1}(t), \ldots, x_{n_1}(t), \ldots, x_{1_n}(t), \ldots, x_{n_n}(t) \right]^T .\]
Due to the exhaustiveness of the system partitioning, the relationship between the compartmental and subcompartmental flows and storages can be stated as
\begin{equation}
\label{eq:cons2_1} 
x_i(t) = \sum \limits_{k=0}^n x_{i_k}(t) \quad \mbox{and} \quad
f_{ij}(t,x) = \sum \limits_{k=0}^n  f_{i_k j_k}(t,{\rm x}) 
\end{equation}
where
\begin{equation}
\label{eq:cons2_1new} 
f_{{i_k}{j_k}}(t,{\rm x}) = x_{j_k}(t) \, \frac{ f_{ij}(t,x) }{x_j(t)} = x_{j_k}(t) \, q^x_{ij}(t,x) = d_{j_k}({\rm x}) \,  f_{ij}(t,x) ,
\end{equation}
and the {\em decomposition factor}, $d_{j_k}({\rm x})$, is defined as $d_{j_k}({\rm x}) = { x_{j_k}(t) }/{x_j(t)}$.

The concepts and notations used in the system partitioning methodology are summarized below:
\begin{center}
    \begin{tabular}[h]{ l p{.70\textwidth} }
    $x_{i_k}(t)$ & storage in subcompartment $k$ of compartment $i$, that is, in subcompartment $i_k$, $k = 0,\ldots,n$, at time $t$, generated by environmental input $z_k(t,{x})$ during $[t_0,t]$ \\ 
    $f_{{i_k}{j_k}}(t,{\rm x})$ & nonnegative flow from subcompartment $j_k$ to $i_k$ at time $t$ \\    
    $y_{i_k}(t,{\rm x}) = f_{0{i_k}}(t,{\rm x})$ & environmental ($j=0$) output from subcompartment $i_k$ at time $t$ \\     
    $z_{i_k}(t,{\rm x}) = \delta_{ik} \, z_{i}(t,x)$ & environmental input into subcompartment $i_k$ at time $t$, where $\delta_{ik}$ is the discrete delta function 
    \\
    \hfill  
    \end{tabular}
\end{center}
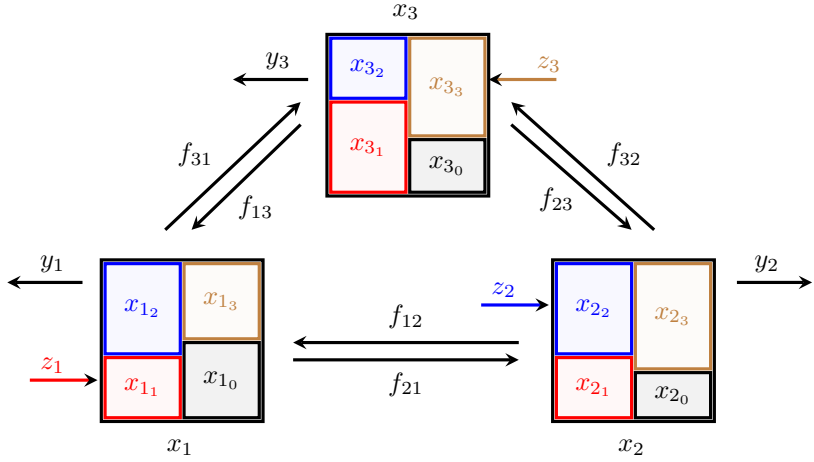
\begin{figure}[t]
\begin{center}
\begin{tikzpicture}
\centering
   \draw[very thick, draw=black] (-.05,-.05) rectangle node(R1) [pos=.5] { } (2.1,2.1) ;
   \draw[very thick, fill=red!3, draw=red, text=red] (0,0) rectangle node(R1) [pos=.5] {$x_{1_1}$} (1,0.8) ;
   \draw[very thick, fill=gray!10, draw=black, text=black] (1.05,0) rectangle node(R2) [pos=.5] {${x}_{1_0}$} (2.05,1) ;
   \draw[very thick, fill=brown!3, draw=brown, text=brown] (1.05,1.05) rectangle node(R3) [pos=.5] {$x_{1_3}$} (2.05,2.05) ;
   \draw[very thick, fill=blue!3, draw=blue, text=blue] (0,0.85) rectangle node(R4) [pos=.5] {$x_{1_2}$} (1,2.05) ;      
    \draw[very thick,-stealth,draw=red]  (-1,.5) -- (-.1,.5) ;     
    \node (z) [text=red] at (-.7,0.7) {$z_1$};                   
    \draw[very thick,-stealth,draw=black]  (-.3,1.8) -- (-1.3,1.8) ;     
    \node (z) [text=black] at (-.7,2.05) {$y_1$};                                                 
    \node (x) at (1,-.4) {${x}_{1}$};                
   \draw[very thick, draw=black] (5.95,-.05) rectangle node(R1) [pos=.5] { } (8.1,2.1) ;
   \draw[very thick, fill=red!3, draw=red, text=red] (6,0) rectangle node(R1) [pos=.5] {$x_{2_1}$} (7,0.8) ;
   \draw[very thick, fill=gray!10, draw=black, text=black] (7.05,0) rectangle node(R2) [pos=.5] {${x}_{2_0}$} (8.05,.6) ;
   \draw[very thick, fill=brown!3, draw=brown, text=brown] (7.05,2.05) rectangle node(R3) [pos=.5] {$x_{2_3}$} (8.05,0.65) ;
   \draw[very thick, fill=blue!3, draw=blue, text=blue] (6,0.85) rectangle node(R4) [pos=.5] {$x_{2_2}$} (7,2.05) ;      
    \draw[very thick,-stealth,draw=blue]  (5,1.5) -- (5.9,1.5) ;     
    \node (z) [text=blue] at (5.3,1.7) {$z_2$};       
    \draw[very thick,-stealth,draw=black]  (8.4,1.8) -- (9.4,1.8) ;     
    \node (z) [text=black] at (8.8,2.05) {$y_2$};                                                 
    \node (x) at (7,-.4) {${x}_{2}$};      
   \draw[very thick, draw=black] (2.95,2.95) rectangle node(R1) [pos=.5] { } (5.1,5.1) ;
   \draw[very thick, fill=red!3, draw=red, text=red] (3,3) rectangle node(R1) [pos=.5] {$x_{3_1}$} (4,4.2) ;
   \draw[very thick, fill=gray!10, draw=black, text=black] (4.05,3) rectangle node(R2) [pos=.5] {${x}_{3_0}$} (5.05,3.7) ;
   \draw[very thick, fill=brown!3, draw=brown, text=brown] (4.05,3.75) rectangle node(R3) [pos=.5] {$x_{3_3}$} (5.05,5.05) ;
   \draw[very thick, fill=blue!3, draw=blue, text=blue] (3,4.25) rectangle node(R4) [pos=.5] {$x_{3_2}$} (4,5.05) ;      
    \draw[very thick,-stealth,draw=brown]  (6,4.5) -- (5.1,4.5) ;     
    \node (z) [text=brown] at (5.9,4.7) {$z_3$};                        
    \draw[very thick,-stealth,draw=black]  (2.7,4.5) -- (1.7,4.5) ;     
    \node (z) [text=black] at (2.3,4.75) {$y_3$};                            
    \node (x) at (4,5.4) {${x}_{3}$};        
    \draw[very thick,-stealth]  (0.8,2.5) -- (2.6,4.2) ;  
    \draw[very thick,-stealth]  (2.6,3.9)  -- (1.15,2.5) ;
    \node (x) at (2,2.8) {${f}_{13}$};  
    \node (x) at (1.2,3.5) {${f}_{31}$};                                           
    \draw[very thick,-stealth]  (7.3,2.5) -- (5.4,4.2) ;  
    \draw[very thick,-stealth]  (5.4,3.9) -- (7,2.5) ;  
    \node (x) at (6,2.9) {${f}_{23}$};  
    \node (x) at (6.9,3.5) {${f}_{32}$};                                           
    \draw[very thick,-stealth]  (2.5,.77) -- (5.5,0.77) ;  
    \draw[very thick,-stealth]  (5.5,1) -- (2.5,1) ;  
    \node (x) at (4,1.35) {${f}_{12}$};  
    \node (x) at (4,.45) {${f}_{21}$};                                           
\end{tikzpicture}
\end{center}
\caption{Schematic representation of the dynamic subcompartmentalization in a three-compartment model system. Each subsystem is colored differently; the second subsystem ($k=2$) is blue, for example. Only the subcompartments in the same subsystem ($x_{1_2}(t)$, $x_{2_2}(t)$, and $x_{3_2}(t)$ in the second subsystem, for example) interact with each other. Subsystem $k$ receives environmental input only at subcompartment ${k_k}$. The initial subsystem receives no environmental input. The dynamic flow partitioning is not represented in this figure. Compare this figure with Fig.~\ref{fig:sc}, in which the subcompartmentalization and corresponding flow partitioning are illustrated for $x_1(t)$ only.}
\label{fig:sd}
\end{figure}

Total subcompartmental inflows and outflows at each compartment $i$ at time $t$ generated by the environmental input into compartment $k$ during $[t_0,t]$ can then be defined, respectively, as follows:
\begin{equation}
\label{eq:in_out_flows_sc}
\check{\tau}_{i_k}(t,{\rm x}) = z_{i_k}(t,{\rm x}) + \sum_{\substack{j=1 }}^n f_{{i_k}{j_k}}(t,{\rm x}) \quad \mbox{and} \quad \hat{\tau}_{i_k}(t,{\rm x}) = y_{i_k}(t,{\rm x}) + \sum_{\substack{j=1 }}^n f_{{j_k}{i_k}}(t,{\rm x})
\end{equation}
for $k=0,1,\ldots,n$. Therefore, $\check{\tau}_{i_k}(t,{\rm x})$ and $\hat{\tau}_{i_k}(t,{\rm x})$ will respectively be called the {\em inward} and {\em outward subthroughflows} at subcompartment $i_k$ at time $t$.

The $n \times n$ {\em inward} and {\em outward subthroughflow} and {\em substorage matrix} functions, $\check{T}(t,{\rm x})$, $\hat{T}(t,{\rm x})$, and $X(t)$, whose entries represent the inward and outward subthroughflows and associated substorages are defined as
\begin{equation}
\label{eq:Tmatrix}
X_{ik}(t) = x_{i_k}(t), \quad 
\check{T}_{ik}(t,{\rm x}) = \check{\tau}_{i_k}(t,{\rm x}), \quad \mbox{and} \quad \hat{T}_{ik}(t,{\rm x}) = \hat{\tau}_{i_k}(t,{\rm x}) 
\end{equation}
for $i,k=1,\ldots,n$. The {\em inward} and {\em outward subthroughflow} and associated {\em substorage vector} functions of size $n$ for the initial subsystem, $\check{\tau}_0(t,{\rm x})$, $\hat{\tau}_0(t,{\rm x})$, and ${x}_0(t)$, are also defined as $\check{\tau}_0(t,{\rm x}) = \left[ \check{\tau}_{1_0}(t,{\rm x}), \ldots, \check{\tau}_{n_0}(t,{\rm x}) \right]^T$, $\hat{\tau}_0(t,{\rm x}) = \left[ \hat{\tau}_{1_0}(t,{\rm x}), \ldots, \hat{\tau}_{n_0}(t,{\rm x}) \right]^T$, and ${x}_0(t) = \left[ x_{1_0}(t), \ldots, x_{n_0}(t) \right]^T$, respectively. The governing equation, Eq.~\ref{eq:model_sc_apx}, can then be expressed in the following compact matrix form: 
\begin{equation}
\label{eq:model_mat}
\begin{aligned}
{\dot X}(t) & = {T}(t,{\rm x}) = \check{T}(t,{\rm x}) - \hat{T}(t,{\rm x}), \quad \, \, \, X (t_0) = \mathbf{0} \\ 
\dot{{x}}_{0}(t) & = {\tau}_{0}(t,{\rm x}) = \check{\tau}_{0}(t,{\rm x}) - \hat{\tau}_{0}(t,{\rm x}), \quad {x}_{0} (t_0) = x_{0} 
\end{aligned} 
\end{equation} 
where $\mathbf{0}$ is used for both $n \times n$ zero matrix and the zero vector of size $n$.

Let the notation $\diag({x(t)})$ represent the diagonal matrix whose diagonal elements are the elements of vector $x(t)$ and $\diag({X(t)})$ represent the diagonal matrix whose diagonal elements are the same as the diagonal elements of matrix $X(t)$. The $n \times n$ diagonal {\em storage}, {\em output}, and {\em input} matrix functions, $\mathcal{X}(t)$, $\mathcal{Y}(t,x)$, and $\mathcal{Z}(t,x)$ will be defined, respectively, as
\[ \mathcal{X}(t) = \operatorname{diag}(x(t)), \quad \mathcal{Y}(t,x) = \operatorname{diag}(y(t,x)), \quad  \mbox{and} \quad \mathcal{Z}(t,x) = \operatorname{diag}(z(t,x)) . \]
Using Eq.~\ref{eq:cons2_1new}, the subthroughflow matrices can then be formulated as follows: 
\begin{equation}
\label{eq:Tmatrix2} 
\begin{aligned}
\check{T}(t,{\rm x}) &= \mathcal{Z}(t,x) + F(t,x) \, \mathcal{X}^{-1}(t) \, X(t) \\ 
\hat{T}(t,{\rm x}) &= \left( \mathcal{Y}(t,x) + \operatorname{diag} \left(F^T(t,x) \, \mathbf{1} \right) \right) \, \mathcal{X}^{-1}(t) \, X(t) \\
&= \mathcal{T}(t,x) \, \mathcal{X}^{-1}(t) \, X(t) \\
\end{aligned}
\end{equation}
where $\mathcal{T}(t,x) = \diag{(\hat{\tau}(t,x))} = \mathcal{Y}(t,x) + \operatorname{diag} \left(F^T(t,x) \, \mathbf{1} \right) $. 

We also define an $n \times n$ matrix function $A(t,x)$ as
\begin{equation}
\label{eq:matrix_A}
\begin{aligned}
A(t,x) &= \left( F (t,x) - \mathcal{Y}(t,x) - \operatorname{diag} \left( F^T (t,x) \, \mathbf{1} \right) \right) \, \mathcal{X}^{-1} (t) \\
& = \left( F(t,x) - \mathcal{T}(t,x) \right) \, \mathcal{X}^{-1} (t) \\
& = Q^x (t,{x}) - \mathcal{R}^{-1}(t,{x}) 
\end{aligned}
\end{equation} 
where $Q^x (t,{x}) = F (t,{x}) \, \mathcal{X}^{-1}(t)$ and $\mathcal{R}^{-1}(t,{x}) = \mathcal{T}(t,{x}) \, \mathcal{X}^{-1}(t)$. Note that the first term in the definition of $A(t,x)$, $Q^x (t,{x})$, represents the intercompartmental flow intensity, and the second term, $\mathcal{R}^{-1}(t,x)$, represents the outward throughflow intensity. Consequently, we call $A(t,x)$ the {\em flow intensity matrix} per unit storage. The $n \times n$ diagonal matrix $\mathcal{R}(t,{x})$ will be called the {\em residence time matrix} and will be discussed further below in Section~\ref{sec:EEtime}. The governing equations, Eq.~\ref{eq:model_mat}, can then be expressed in the following form:
\begin{equation}
\label{eq:model_M}
\begin{aligned}
\dot{X}(t) & = \mathcal{Z} (t,x) + A(t,x) \, X(t) ,
\quad X(t_0) = \mathbf{0} , \\
\dot{{x}}_{0}(t) & = A(t,x) \, x_0(t) ,
\quad \quad \quad \quad \quad  {x}_{0} (t_0) = x_{0} ,
\end{aligned}
\end{equation} 
as formulated by \cite{Coskun2017DCSAM}.

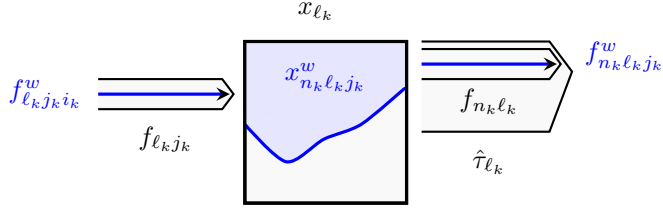
\begin{figure}[t]
\begin{center}
\begin{tikzpicture} 
   \draw[very thick, fill=gray!5, draw=black] (-.05,-.05) rectangle node(R1) [pos=.5] { } (2.1,2.1) ;
   \draw [very thick, fill=blue!10, draw=blue] plot [smooth] coordinates {(-0.05,1) (.5,.5) (1,.8) (1.5,1) (2.1,1.5)};
   \draw [fill=blue!10, draw=none]  (-0.05,1) -- (2.1,1.5) -- (2.1,2.1) -- (-.05,2.1) ;   
   \draw[very thick, draw=black, text=blue] (-.05,-.05) rectangle node(R1) [pos=.5, yshift=.6cm] { $x^w_{n_k \ell_k j_k}$ } (2.1,2.1) ;   
   \draw [thick, fill=gray!5, draw=black]  (-2,1.2) -- (-0.35,1.2) -- (-0.2,1.4) -- (-.35,1.6) -- (-2,1.6) ;      
   \draw [thick, fill=gray!5, draw=black]  (2.3,.9) -- (4,.9) -- (4.3,1.7) -- (4,2.1) -- (2.3,2.1) ;           
   \draw [thick, fill=gray!5, draw=black]  (2.3,1.6) -- (4,1.6) -- (4.15,1.8) -- (4,2) -- (2.3,2) ;
    \node (x) at (.9,2.5) {${x}_{\ell_k}$};  
    \node (x) at (-1.1,.8) {$f_{\ell_k j_k}$};
   \draw [very thick, -stealth, draw=blue]  (-2,1.4) -- (-.25,1.4) ;
    \node [text=blue] (x) at (-2.7,1.4) {$f^w_{\ell_k j_k i_k}$};
   \draw [very thick, -stealth, draw=blue]  (2.3,1.8) -- (4.1,1.8) ; 
    \node [text=blue] (x) at (5,1.9) {$f^w_{n_k \ell_k j_k}$}; 
    \node (x) at (3.2,1.3) {$f_{n_k \ell_k}$};   
    \node (x) at (3.2,.5) {${\hat{\tau}}_{\ell_k}$};    
\end{tikzpicture}
\end{center}
\caption{Schematic representation of the dynamic subsystem decomposition. The transient inflow and outflow functions, $f^w_{\ell_k j_k i_k}(t)$ and $f^w_{n_k \ell_k j_k}(t)$, at and associated transient substorage, $x^w_{n_k \ell_k j_k}(t)$, in subcompartment ${\ell_k}$ along subflow path $p^w_{n_k j_k}= i_k \mapsto j_k \to \ell_k \to n_k$. }
\label{fig:subsystemp}
\end{figure}

The transient flows and storages have also been recently introduced by \cite{Coskun2017DCSAM,Coskun2017NDP}. The {\em transient subflows} along a subflow path is defined through the {\em subsystem partitioning methodology} as follows: Along a given subflow path $p^w_{n_k j_k}= i_k \mapsto j_k \to \ell_k \to n_k$, the {\em transient inflow} at subcompartment ${\ell_k}$, $f^{w}_{\ell_k j_k i_k}(t)$, generated by the local input from $i_k$ to ${j_k}$ during $[t_1,t]$, $t_1 \geq t_0$, is the input segment that is transmitted from $j_k$ to ${\ell_k}$ at time $t$. Similarly, the {\em transient outflow} generated by the transient inflow at ${\ell_k}$ during $[t_1,t]$, $f^{w}_{n_k \ell_k j_k}(t)$, is the inflow segment that is transmitted from ${\ell_k}$ to the next subcompartment, ${n_k}$, along the path at time $t$. The associated {\em transient substorage} in subcompartment ${\ell_k}$ at time $t$, $x^{w}_{n_k \ell_k j_k}(t)$, is the substorage segment governed by the transient inflow and outflow balance during $[t_1,t]$  (see Fig.~\ref{fig:subsystemp}).

The transient outflow at subcompartment ${\ell_k}$ at time $t$ along subflow path $p^w_{n_k j_k}$ from ${j_k}$ to ${n_k}$, $f^{w}_{n_k \ell_k j_k}(t)$, can be formulated as follows:
\begin{equation}
\label{eq:out_in_fs}
\begin{aligned}
f^{w}_{n_k \ell_k j_k}(t) = \frac{ f_{n_k \ell_k}(t,\rm{x}) }{x_{\ell_k}(t) } \, x^{w}_{n_k \ell_k j_k}(t) ,
\end{aligned}
\end{equation}
due to the equivalence of flow and subflow intensities, where the transient substorage, $x^{w}_{n_k \ell_k j_k}(t)$, is determined by the governing mass balance equation 
\begin{equation}
\label{eq:out_in_fs2}
\begin{aligned}
\dot{x}^{w}_{n_k \ell_k j_k}(t) & = f^{w}_{\ell_k j_k i_k}(t) - \frac{ {\hat{\tau}}_{\ell_k}(t,\rm{x}) }{ x_{\ell_k}(t) } \, {x}^{w}_{n_k \ell_k j_k}(t) , \quad {x}^{w}_{n_k \ell_k j_k}(t_1) = 0 .
\end{aligned}
\end{equation}
The equivalence of the throughflow and subthroughflow intensities, as well as the flow and subflow intensities in the same direction, that is
\begin{equation}
\label{eq:eqiv_intense}
\begin{aligned}
q^x_{n \ell}(t,x) = \frac{  f_{n \ell}(t,x) }{x_{\ell}(t) } = \frac{ f_{n_k \ell_k}(t,\rm{x})  }{x_{\ell_k}(t) } 
 \quad \mbox{and} \quad r^{-1}_\ell (t,x) = \frac{ {\hat{\tau}}_{\ell}(t,x) }{ x_{\ell}(t) } = \frac{ {\hat{\tau}}_{\ell_k}(t,\rm{x}) }{ x_{\ell_k}(t) }
 \end{aligned}
\end{equation}
are given by Eqs.~\ref{eq:cons2_1new} and~\ref{eq:Tmatrix2}, for $\ell,n=1,\ldots,n$, and $k=0,1,\ldots,n$ \cite{Coskun2017NDP}. Therefore, since the intensities in Eqs.~\ref{eq:out_in_fs} and~\ref{eq:out_in_fs2} can be expressed at both the subcompartmental and compartmental levels, the transient flows and storages can be determined along both subflow paths within the subsystems and flow paths within the system. This allows the flexibility of tracking arbitrary intercompartmental flows and storages generated by all or individual environmental inputs within the system.

\begin{figure}[t]
\begin{center}
\begin{tikzpicture}
   \draw[very thick, fill=gray!10, draw=black] (-.05,-.05) rectangle node(R1) [pos=.5] { } (2.1,2.4) ;
   \draw[very thick, fill=blue!3, draw=blue, text=blue] (0.3,0.1) rectangle node(R1) [pos=.5] {$x_{i_i}$} (1.2,1.3) ;
    \node (x) at (1,-.5) {${x}_{i}$};
   \draw[very thick, fill=gray!10, draw=black] (8.95,-.05) rectangle node(R2) [pos=.5] { } (11.1,2.4) ;
   \draw[very thick, fill=gray!10, draw=black] (8.95,1.69) rectangle node(R2) [pos=.5] { } (10.5,2.3) ;   
   \draw[very thick, fill=blue!3, draw=blue, text=blue] (9.05,0.05) rectangle node(R1) [pos=.5] {$x_{j_i}$} (10.05,1.1) ;
    \node (x) at (10,-.5) {${x}_{j}$};
    \node (x) at (9.7,2) {${x}_{j_0}$};    
   \draw[fill=gray!10] (3.05,1.3) -- ++ (-0.6,0) -- ++ (-0.3,.3) -- ++ (.3,.3) -- ++ (.6,0);
   \draw[fill=gray!10] (8.1,1.3) -- ++ (0.84,0) -- ++ (0,.4) -- ++ (-.84,0) ;
   \draw[fill=gray!10] (8.1,1.7) -- ++ (0.84,0) -- ++ (0,.43) -- ++ (-.84,0) ;
    \node[] (t2) at (2.8,1) {${\tau}^\texttt{t}_{i j}$};
    \node[] (t3) at (8.5,2.4) {$\hat{\tau}_{j}$};
    \node[] (t3) at (8.5,1.92) {$\hat{\tau}_{j_0}$};    
    \node[] (t2) at (4,2.2) {${\tau}^\texttt{c}_{i j}$};
    \node[] (x) at (4.5,1.1) {${\tau}^\texttt{d}_{i j} $};
    \draw[very thick,-stealth]  (8.8,1.4) -- (2.4,1.4);
   \draw[fill=blue!10] (2.95,0.3) -- ++ (-0.84,0) -- ++ (0,.4) -- ++ (.84,0) ;
    \draw[very thick,-stealth,draw=blue]  (-.8,.5) -- (.25,.5) ;
    \node (z) [text=blue] at (-.45,.8) {$z_i$};
    \node[blue] (t3) at (2.7,0.05) {$\hat{\tau}_{i_i}$};
   \draw[fill=blue!10] (8,0.1) -- ++ (0.6,0) -- ++ (0.3,.3) -- ++ (-.3,.3) -- ++ (-.6,0);
    \node[blue] (t1) at (8.4,1) {$\tilde{\tau}_{j_i}$};
    \draw[very thick,draw=blue, -stealth]  (1.3,0.6) -- (8.7,0.6) ;
    \node[blue] (x) at (6.7,0.9) {${\tau}^\texttt{d}_{j_i} $};
   \node[blue,anchor=west] at (3.3,0.1) (a) {${\tau}^\texttt{i}_{j_i}$};
   \node[blue,anchor=east] at (7,0.4) (b) {};
   \node[blue,anchor=west] at (.9,0.4) (e) {};
   \node[blue,anchor=east] at (3.3,0.1) (f) {};
\draw [very thick,draw=blue, dashed, -stealth] plot [smooth, tension=1] coordinates { (1.3,0.4) (4,0.4) (5,-0.1) (4.5,-0.6) (4,-0.1) (5,0.4) (8.7,0.4) };
\draw [very thick,draw=blue, dashed, -stealth] plot [smooth, tension=1] coordinates { (9,0.4) (9.6,-0.1) (9,-0.5) (7,-0.5) (6.5,-0.1) (7,0.2) (8.7,0.2) };
    \node[blue] (t2) at (7,-0.2) {${\tau}^\texttt{c}_{j_i}$};
   \node[anchor=east] at (7.2,2.1) (c) {$\tau^\texttt{i}_{i j}$};
   \node[blue,anchor=west] at (2.1,1.7) (d) {};
\draw [very thick,dashed, -stealth] plot [smooth, tension=1] coordinates { (8.8,1.6) (6.5,1.6)  (5.5,2.1)  (6,2.6) (6.5,2.1) (5.5,1.6) (2.4,1.6) };
\draw [very thick, dashed, -stealth] plot [smooth, tension=1] coordinates { (2,1.6) (1.5,2.1) (2,2.6) (4,2.6) (4.5,2.1) (4,1.8) (2.4,1.8) };
   \node[blue,anchor=east] at (5.9,2.1) (h) {};
   \node[blue,anchor=west] at (8,1.7) (g) {};
\end{tikzpicture}
\end{center}
\caption{Schematic representation of the simple and composite \texttt{diact} flows. Solid arrows represent direct flows, and dashed arrows represent indirect flows through other compartments (not shown).
The composite \texttt{diact} flows (black) generated by outward throughflow $\hat{\tau}_{j}(t,{x}) - \hat{\tau}_{j_0}(t,{\rm x})$ (i.e. derived from all environmental inputs): direct flow, $\tau^\texttt{d}_{i j}(t)$, indirect flow, $\tau^\texttt{i}_{ij}(t)$, acyclic flow, $\tau^\texttt{a}_{ij}(t) = \tau^\texttt{t}_{ij}(t) - \tau^\texttt{c}_{ij}(t)$, cycling flow, $\tau^\texttt{c}_{ij}(t)$, and transfer flow, $\tau^\texttt{t}_{ij}(t)$.
The simple \texttt{diact} flows (blue) generated by outward subthroughflow $\hat{\tau}_{i_i}(t,{\rm x})$ (i.e.  derived from single environmental input $z_{i}(t)$): direct flow, ${\tau}^\texttt{d}_{j_i}(t) = {\tau}^\texttt{d}_{j_i i_i}(t) $, indirect flow, ${\tau}^\texttt{i}_{j_i}(t) = \tau^\texttt{i}_{j_i i_i}(t)$, acyclic flow, ${\tau}^\texttt{a}_{j_i}(t) = \tau^\texttt{a}_{j_i i_i}(t) = {\tau}^\texttt{t}_{j_i}(t) - {\tau}^\texttt{c}_{j_i}(t)$, cycling flow, ${\tau}^\texttt{c}_{j_i}(t) = \tau^\texttt{c}_{j_i i_i}(t)$, and transfer flow, $\tau^{\texttt{t}}_{j_i}(t) = \tau^{\texttt{t}}_{j_i i_i}(t) = \tilde{\tau}_{j_i}(t,{\rm x}) = \check{\tau}_{j_i}(t,{\rm x}) - z_{j_i}(t)$. Note that the cycling flows at the terminal (sub)compartment may include the segments of the direct and/or indirect flows at that (sub)compartment, if the cycling flows indirectly pass through the corresponding initial (sub)compartment. Therefore, the acyclic flows are composed of the segments of the direct and/or indirect flows.
}
\label{fig:utilityfigs}
\end{figure}
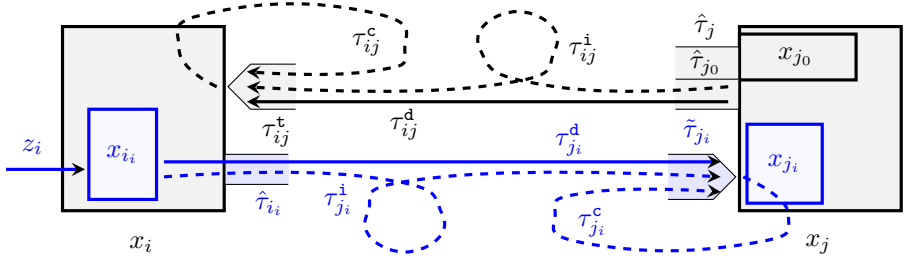

The \texttt{diact} flows and storages have also been recently formulated through the {\em dynamic} and {\em path-based approaches} based on the system and subsystem partitioning methodologies, respectively, by \cite{Coskun2017DCSAM,Coskun2017NDP}. The {\em composite transfer flow} is defined as the total intercompartmental transient flow that is generated by all environmental inputs from one compartment, {\em directly} or {\em indirectly} through other compartments, to another. The {\em composite direct}, {\em indirect}, {\em acyclic}, and {\em cycling flows} from the initial compartment to the terminal compartment are then defined as the direct, indirect, non-cycling, and cycling segments at the terminal compartment of the composite transfer flow (see Fig.~\ref{fig:utilityfigs}). The cycling and acyclic flows can be interpreted as the flows that visit the terminal compartment {\em multiple times} and {\em only once}, respectively, after being transmitted from the initial compartment. 

The {\em composite transfer subflow} within the initial subsystem can also be defined as the total intercompartmental transient flow, derived from all initial stocks, from one initial subcompartment {\em directly} or {\em indirectly} through other initial subcompartments to another. The {\em composite direct}, {\em indirect}, {\em acyclic}, and {\em cycling subflows} within the initial subsystem from the initial subcompartment to the terminal subcompartment are then defined as the direct, indirect, non-cycling, and cycling segments at the terminal subcompartment of the composite transfer subflow.

The {\em simple transfer flow} will be defined as the total intercompartmental transient subflow that is generated by the single environmental input from an input-receiving subcompartment, {\em directly} or {\em indirectly} through other compartments, to another subcompartment. The {\em simple direct}, {\em indirect}, {\em acyclic}, and {\em cycling flows} from the initial input-receiving subcompartment to the terminal subcompartment are then defined as the direct, indirect, non-cycling, and cycling segments at the terminal subcompartment of the simple transfer flow (see Fig.~\ref{fig:utilityfigs}). The associated simple and composite \texttt{diact} storages are defined as the storages generated by the corresponding \texttt{diact} flows.

The composite \texttt{diact} subflows from subcompartment $k_\ell$ to $i_\ell$ at time $t$ are formulated through the dynamic approach as follows:
\begin{equation}
\label{eq:comp_diact_subs}
\begin{aligned}
 \tau^{\texttt{d}}_{i_\ell k_\ell}(t) & =  \frac{ f_{i_k k_k}(t,{\rm x}) } {{\hat{\tau}}_{k_k}(t,{\rm x}) } \, {{\hat{\tau}}_{k_\ell}(t,{\rm x}) } = \frac{ f_{i k}(t,{x}) } {{\hat{\tau}}_{k}(t,{x}) } \, {{\hat{\tau}}_{k_\ell}(t,{\rm x}) } \\ 
 \tau^{\texttt{i}}_{i_\ell k_\ell}(t) & = \frac { {\check{\tau}}_{i_k} (t,{\rm x}) - z_{i_k}(t,{\rm x}) - f_{i_k k_k}(t,{\rm x}) } { {\hat{\tau}}_{k_k}(t,{\rm x}) } \, { {\hat{\tau}}_{k_\ell}(t,{\rm x}) } \\
 \tau^{\texttt{a}}_{i_\ell k_\ell}(t) & = \left[ \frac{ {\check{\tau}}_{i_k} (t,{\rm x}) - z_{i_k}(t,{\rm x}) } {{\hat{\tau}}_{k_k}(t,{\rm x}) } - \frac{ {\check{\tau}}_{i_i}(t,{\rm x}) - z_{i_i}(t,{\rm x}) } {{\hat{\tau}}_{i_i}(t,{\rm x}) } \, \frac{ {\hat{\tau}}_{i_k}(t,{\rm x}) } {{\hat{\tau}}_{k_k}(t,{\rm x}) }  \right] {{\hat{\tau}}_{k_\ell}(t,{\rm x}) } \\
 \tau^{\texttt{c}}_{i_\ell k_\ell}(t) & = \frac{ {\check{\tau}}_{i_i}(t,{\rm x}) - z_{i_i}(t,{\rm x}) } {{\hat{\tau}}_{i_i}(t,{\rm x}) } \, \frac{ {\hat{\tau}}_{i_k}(t,{\rm x}) } {{\hat{\tau}}_{k_k}(t,{\rm x}) }  \, {{\hat{\tau}}_{k_\ell}(t,{\rm x}) } \\
 \tau^{\texttt{t}}_{i_\ell k_\ell}(t) & = \frac{ {\check{\tau}}_{i_k} (t,{\rm x}) - z_{i_k}(t,{\rm x}) }  {{\hat{\tau}}_{k_k}(t,{\rm x}) } \, { {\hat{\tau}}_{k_\ell}(t,{\rm x}) }
\end{aligned}
\end{equation}
for $t > t_0$, $i,k=1,\ldots,n$, and $\ell=0,\ldots,n$, using the proportionality of parallel subflows \cite{Coskun2017DCSAM,Coskun2017NDP}. Note that $\hat{\tau}_{k_k}(t_0,{\rm x})=0$ and we assume that $\hat{\tau}_{k_k}(t,{\rm x})$ is nonzero for all $t > t_0$. The simple and composite \texttt{diact} flow matrices are listed in matrix form in Table~\ref{tab:flow_stor1}. We use a tilde notation over the simple versions of the \texttt{diact} flow vector and matrix quantities. The diagonal matrices $\check{\mathsf{T}}(t,{\rm x})$, $\hat{\mathsf{T}}(t,{\rm x})$, and $\tilde{\mathsf{T}}(t,{\rm x})$ used in Table~\ref{tab:flow_stor1} are defined as
\begin{equation}
\label{eq:two_matx}
\begin{aligned}
\check{\mathsf{T}}(t,{\rm x}) = \diag{( \check{T}(t,{\rm x}) )} , \, \, \,
\hat{\mathsf{T}}(t,{\rm x}) = \diag{( \hat{T}(t,{\rm x}) )} , \, \, \,
\tilde{\mathsf{T}}(t,{\rm x}) = \diag{( \tilde{T}(t,{\rm x}) )} ,
\end{aligned}
\end{equation}
where $\tilde{T}(t,{\rm x}) = \check{T}(t,{\rm x}) - \mathcal{Z}(t,x)$. The inverted matrices in the table are assumed to be invertible.
\begin{table}
     \centering
     \caption{The dynamic \texttt{diact} flow distribution and the simple and composite \texttt{diact} (sub)flow matrices. The superscript ($^\texttt{*}$) in each equation represents any of the \texttt{diact} symbols. For the sake of readability, the function arguments are dropped.}
     \label{tab:flow_stor1}
     \begin{tabular}{c p{6cm} l }
     \hline
\texttt{diact} & {flow distribution matrix} & {flows} \\
     \hline
     \noalign{\vskip 2pt}
\texttt{d} & $ N^\texttt{d} =  F \, \mathcal{T}^{-1}  $  &
\multirowcell{5}{ 
$
\begin{aligned}
\hfill
{T}^\texttt{*} &= {N}^\texttt{*} \, (\mathcal{T} -  \mathcal{\hat T}_0) \\ 
{T}^\texttt{*}_\ell &= {N}^\texttt{*} \, \mathcal{\hat T}_\ell \\ 
\tilde{T}^\texttt{*} &= {N}^\texttt{*} \, \mathsf{\hat T}  
\end{aligned}
$
} 
\\
\texttt{i} & $ N^\texttt{i} = \displaystyle \tilde{T} \, \hat{\mathsf{T}}^{-1} - F \,\mathcal{T}^{-1}  $ &  \\
\texttt{a} & $ {N}^\texttt{a} = \displaystyle \tilde{T} \, \hat{\mathsf{T}}^{-1} - \tilde{\mathsf{T}} \, \hat{\mathsf{T}}^{-1} \, \hat{T} \, \hat{\mathsf{T}}^{-1} $ & \\
\texttt{c} & $ {N}^\texttt{c} = \displaystyle \tilde{\mathsf{T}} \, \hat{\mathsf{T}}^{-1} \, \hat{T} \, \hat{\mathsf{T}}^{-1} $
& \\
\texttt{t} & $ N^\texttt{t} = \displaystyle  \tilde{T} \, \hat{\mathsf{T}}^{-1}   $ & \\
\noalign{\vskip 1pt}
\hline
     \end{tabular}
\end{table}

The simple and composite \texttt{diact} flows, $\tau^{\texttt{*}}_{i_k}(t)$ and $\tau^{\texttt{*}}_{i k}(t)$, and storages, $x^{\texttt{*}}_{i_k}(t)$ and $x^{\texttt{*}}_{i k}(t)$, generated by environmental inputs can then be defined in terms of the composite \texttt{diact} subflows and substorages as follows:
\begin{equation}
\label{eq:simple_comp}
\begin{aligned}
\tau^{\texttt{*}}_{i_k}(t) = \tau^{\texttt{*}}_{i_k k_k}(t) 
\quad & \mbox{and} \quad 
{\tau}^{\texttt{*}}_{ik}(t) = \sum_{\ell=1}^{n} {\tau}^\texttt{*}_{i_\ell k_\ell} (t) , \\
{x}^{\texttt{*}}_{i_k}(t) = {x}^{\texttt{*}}_{i_k k_k}(t) 
\quad & \mbox{and} \quad 
x^{\texttt{*}}_{ik}(t) = \sum_{\ell=1}^{n} x^\texttt{*}_{i_\ell k_\ell} (t) .
\end{aligned}
\end{equation}
Here, the composite \texttt{diact} substorages are formulated as
\begin{equation}
\label{eq:out_in_diact2}
\begin{aligned}
\dot{x}^{\texttt{*}}_{i_\ell k_\ell}(t) & = \tau^{\texttt{*}}_{i_\ell k_\ell}(t) - \frac{ {\hat{\tau}}_{i}(t,{x}) }{ x_{i}(t) } \, {x}^{\texttt{*}}_{i_\ell k_\ell}(t) , \quad {x}^{\texttt{*}}_{i_\ell k_\ell}(t_1) = 0
\end{aligned}
\end{equation}
for $t_1 > t_0$, $i,k=1,\ldots,n$, and $\ell=0,\ldots,n$ \cite{Coskun2017DCSAM}. The solution to this governing equation, ${x}^{\texttt{*}}_{i_\ell k_\ell}(t)$, represents the \texttt{diact} substorage at time $t \geq t_1$ generated by the corresponding \texttt{diact} subflow, $\tau^{\texttt{*}}_{i_\ell k_\ell}(t)$, during $[t_1,t]$.

The proposed methodology constructs a base for the development of new mathematical system analysis tools. Multiple dynamic {\em measures} and {\em indices} of matrix, vector, and scalar types are developed as quantitative ecological indicators in the present paper. Since the dynamic measures are functions of time, their time derivatives and integrals can also be used for further analysis of various system attributes as formulated in what follows.

We will start with a brief summary and interpretations of the measures developed in this section, such as the substorages, subthroughflows, as well as the transient and dynamic \texttt{diact} flows and storages. The systematic formulation of new mathematical system analysis tools will follow that discussion. The static versions of these measures and indices has recently been formulated in a separate article~\cite{Coskun2017SESM}.

\subsection{System measures}
\label{sec:stss}

The dynamic system partitioning methodology yields the subthroughflow and substorage matrices that measure the environmental influence on system compartments in terms of the flow and storage generation. For the quantification of intercompartmental flow and storage dynamics, the transient and $\texttt{diact}$ flows and associated storages are formulated through the system and subsystem partitioning methodologies.

The elements of the net subthroughflow and substorage matrices, ${T}(t,{\rm x})$ and $X(t)$, represent the distribution of environmental inputs and the organization of the associated storages generated by the inputs within the system. More precisely, ${\tau}_{i_k}(t,{\rm x})$ and $x_{i_k}(t)$ represent the net subthroughflow at and substorage in compartment $i$ at time $t$ generated by the environmental input into compartment $k$, $z_k(t)$, during $[t_0,t]$ (see Fig.~\ref{fig:sd} and~\ref{fig:sc}). In other words, the proposed methodology can dynamically partition composite compartmental throughflows and associated storages into subcompartmental segments based on their constituent sources from environmental inputs of the same conserved quantity. This partitioning enables tracking the evolution of environmental inputs and the associated storages generated by the inputs individually and separately within the system. Note that the composite compartmental net throughflow and storage, ${\tau}_{i}(t,x)$ and $x_i(t)$, cannot be used to distinguish the portions of this throughflow and storage derived from individual environmental inputs separately. Therefore, the solution to the decomposed system brings out inferences that cannot be obtained through the analysis of the original system by the state-of-the-art techniques. The arguments presented for the net subthroughflow functions above are also valid separately for the inward and outward subthroughflow functions, $\check{\tau}_{i_k}(t,{\rm x})$ and $\check{\tau}_{i_k}(t,{\rm x})$, as well. Similarly, the initial substorage and subthroughflow vectors, $x_0(t)$ and ${\tau}_0(t,{\rm x})$, represent the organization of the initial stocks and the distribution of the associated flows emanating from these stocks within the system.

The {\em transient} flows and associated storages transmitted along a given subflow path are also formulated through the dynamic subsystem partitioning methodology. Therefore, the dynamic {\em subsystem} partitioning determines the distribution of arbitrary intercompartmental flows and the organization of the associated storages generated by these flows along any given subflow path within the subsystems. Consequently, arbitrary composite intercompartmental flows and storages can dynamically be decomposed into the constituent transient subflow and substorage segments along a given set of subflow paths. In other words, the subsystem decomposition enables dynamically tracking the fate of arbitrary intercompartmental flows and associated storages within the subsystems. 

The dynamic direct, indirect, acyclic, cycling, and transfer (\texttt{diact}) flows and storages transmitted from one compartment directly or indirectly through other compartments to any other\textemdash including itself\textemdash within the system are also systematically formulated to determine the intercompartmental flow and storage dynamics.

\subsection{The \texttt{diact} effect measures and indices}
\label{sec:effects}

The effect of one compartment on another through direct transactions is relatively easier to analyze, even in complex networks. The proposed subsystem partitioning methodology enables also the determination of the effect of one compartment indirectly through other compartments on another or itself within the system. In fact, parallel to the definitions of the direct, indirect, acyclic, cycling and transfer (\texttt{diact}) flows and storages, we systematically introduce all the \texttt{diact} effect measures and indices in this section.

Based on the transfer (or total) flows introduced above, the {\em transfer (total) effect measures} and {\em indices} and the corresponding {\em system efficiencies} and {\em stress} are formulated below at the compartmental level. The subcompartmental level formulations in parallel are straightforward, using the transfer subflows and substorages instead of the transfer flows and storages. 

The {\em flow-based transfer effect index} will be defined as the transfer flow normalized by total system throughflow. The flow-based {\em system transfer efficiency} will then be defined as the time derivative of the effect index, and, so, it measures the rate of change of the dynamic index. The {\em storage-based transfer effect index} and {\em system efficiency} can similarly be defined, using the transfer storages instead of the transfer flows and total system storage for normalization. Therefore, the flow- and storage-based effect indices are fractions of total inward system throughflow and storage, respectively. Both flow- and storage-based effect indices quantify the influence of system compartments on each other. The storage-based formulations represent the history of interactions during $[t_1,t]$ while the flow-based formulations represent simultaneous interactions at time $t$.

The {\em flow-} and {\em storage-based composite transfer effect indices}, $\texttt{t}_{IK}^\tau(t)$ and $\texttt{t}_{IK}^x(t)$, of a set of compartments, $K$, on another set, $I$, induced by environmental inputs can be formulated as the fraction of total inward system throughflow that is initiated at compartments $K$ during $[t_1,t]$, $t_1 \geq t_0$ and transmitted {\em directly or indirectly} to $I$ at time $t$, and as the fraction of total system storage generated by these transfer flows during $[t_1,t]$, respectively. That is, 
\begin{equation}
\label{eq:e_wc4}
\begin{aligned}
\texttt{t}^\tau_{IK}(t) = \frac{ \sum_{\substack{i \in I }} \sum_{\substack{k \in K }}  \tau^\texttt{t}_{i k}(t) }{ \sum_{\substack{i=1}}^n \check{\tau}_{i}(t) } \quad \mbox{and} \quad 
\texttt{t}^x_{IK}(t) = \frac{ \sum_{\substack{i \in I }} \sum_{\substack{k \in K }} x^\texttt{t}_{i k}(t) }{  \sum_{\substack{i=1}}^n x_{i}(t) } 
\end{aligned}
\end{equation}
where $I,K \subseteq \{ 1,\ldots,n \}$. If the sets $I$ and $K$ have one element, that is, $I=\{ i\}$ and $K=\{ k\}$, the transfer effect indices will be denoted by $\texttt{t}^\tau_{ik}(t)$ and $\texttt{t}^x_{ik}(t)$. Note that the transfer effect indices are dimensionless. The flow- and storage-based composite transfer effect indices of initial subcompartment $k_0$ on $i_0$ induced by the initial stocks, $\texttt{t}_{i_0 k_0}^\tau(t)$ and $\texttt{t}_{i_0 k_0}^x(t)$, can be formulated similarly by using  the corresponding composite transfer subflows and substorages, $\tau^\texttt{t}_{i_0 k_0}$ and $x^\texttt{t}_{i_0 k_0}$, respectively.

The {\em flow-} and {\em storage-based composite transfer effect matrix} measures induced by environmental inputs will be denoted by $\texttt{T}^\tau(t) = \left( \texttt{t}^\tau_{ik}(t) \right)$ and $\texttt{T}^x(t) = \left( \texttt{t}^x_{ik}(t) \right)$ and formulated in matrix form as
\begin{equation}
\label{eq:b_utilityU2}
\begin{aligned}
\texttt{T}^\tau(t) = 
\frac{ 1 } { \check{\sigma}^\tau(t) } \, {T}^{\texttt{t}}(t) \quad \mbox{and} \quad 
\texttt{T}^x(t) = \frac{ 1 } { {\sigma}^x(t) } \, {X}^{\texttt{t}}(t)
\end{aligned}
\end{equation}
where the {\em scalar} functions $\check{\sigma}^{\tau}(t) = \bm{1}^T \, \check{\tau}(t,x)$, $\hat{\sigma}^{\tau}(t) = \bm{1}^T \, \hat{\tau}(t,x)$, and ${\sigma}^{x}(t) = \bm{1}^T \, x(t)$ are the {\em total inward}, {\em outward system throughflow} and {\em system storage}, respectively. The {\em flow-based composite transfer effect} of the system on the compartments, $\check{\texttt{t}}^\tau(t)$, and those of the compartments on the system, $\hat{\texttt{t}}^\tau(t)$, induced by environmental inputs will be defined as {\em vector measures}:
\begin{equation}
\label{eq:b_utilityU3}
\begin{aligned}
\check{\texttt{t}}^\tau(t) = \texttt{T}^\tau(t) \, \mathbf{1} \quad \mbox{and} \quad \hat{\texttt{t}}^\tau(t) = \mathbf{1}^T \, \texttt{T}^\tau(t).
\end{aligned}
\end{equation}
The {\em storage-based composite transfer effect vectors} can be defined similarly. We will use the notations $\texttt{t}^\tau(t)$ and $\texttt{t}^x(t)$ for the sum of the transfer effects of the entire system on all compartments, that is, for $I=K=\{1,\ldots,n\}$. They can be formulated as
\begin{equation}
\label{eq:b_effT}
\begin{aligned}
\texttt{t}^\tau(t) & 
= \frac{ \mathbf{1}^T \, {T}^{\texttt{t}} (t) \,  \bm{1} } { \check{\sigma}^{\tau}(t) } 
=  \mathbf{1}^T \, \texttt{T}^{\tau}(t)  \,  \mathbf{1} 
\quad \mbox{and} \quad
\texttt{t}^x(t) 
= \frac{ \mathbf{1}^T \, {X}^{\texttt{t}} (t) \,  \bm{1} } { \sigma^{x}(t) } 
=  \mathbf{1}^T \, \texttt{T}^x(t)  \,  \mathbf{1} .
\end{aligned}
\end{equation}
These {\em scalar} functions will be called the {\em flow-} and {\em storage-based composite transfer effect indices} for the system induced by environmental inputs.

The dynamic measures are functions of time, and their time derivatives and integrals also represent various system attributes. In addition to the local-in-time indices introduced above, the {\em average} or {\em non-local composite transfer effect indices} induced by environmental inputs over time interval $[t_1,t]$ can be defined by integrating both the numerators and denominators of $\texttt{t}^\tau(t)$ and $\texttt{t}^x(t)$ separately over the interval. That is,
\begin{equation}
\label{eq:cindex_intT}
\texttt{t}^\tau(t_1,t) = 
\frac{ \int_{t_1}^{t} \mathbf{1}^T \, {T}^{\texttt{t}} (s) \,  \bm{1} \, ds } { \int_{t_1}^{t} \check{\sigma}^{\tau}(s) \, ds } 
\quad \mbox{and} \quad 
\texttt{t}^x(t_1,t) = \frac{ \int_{t_1}^{t} \mathbf{1}^T \, {X}^{\texttt{t}} (s) \,  \bm{1} \, ds } { \int_{t_1}^{t} \sigma^{x}(s) \, ds }  .
\end{equation}
The integrals of the transfer flows and storages involved in the formulations above, $\int_{t_1}^t {\tau}^\texttt{t}_{i j}(s) \, ds$ and $\int_{t_1}^t { x}^\texttt{t}_{i j}(s) \, ds$, measure the total composite transfer flows transmitted and associated storages generated during $[t_1,t]$, respectively. Similarly, $\int_{t_1}^t \check{\sigma}^{\tau}(s) \, ds$ and $\int_{t_1}^t \sigma^{x}(s) \, ds$ are the {\em cumulative} total system throughflow and storage during the same time period.

The time derivatives of the composite transfer effect indices, $\dot{\texttt{t}}^\tau(t)$ and $\dot{\texttt{t}}^x(t)$, will be called the {\em composite transfer flow} and {\em storage efficiencies} for the system induced by the environmental inputs, respectively, as the higher rates indicate increasing transfer effects and, consequently, more {\em efficient} compartmental transactions. They can be formulated as
\begin{equation}
\label{eq:cirT}
\dot{\texttt{t}}^\tau(t) = \frac{d}{dt} \left( \frac{ \mathbf{1}^T \, {T}^{\texttt{t}} (t) \,  \bm{1} } { \check{\sigma}^{\tau}(t) }  \right) \quad \mbox{and} \quad \dot{\texttt{t}}^x(t) = \frac{d}{dt} \left( \frac{ \mathbf{1}^T \, {X}^{\texttt{t}} (t) \,  \bm{1} } { \sigma^{x}(t) }  \right) .
\end{equation}
The time derivatives of the transfer flows and storages involved in the formulations above, $\dot{\tau}^\texttt{t}_{i j}(t)$ and $\dot{x}^\texttt{t}_{i j}(t)$, measure the rate of change of the composite transfer flows and storages at time $t$. Similarly, $\dot{\check{\sigma}}^{\tau}(t)$ and $\dot{\sigma}^{x}(t)$ are the rate of change of the total system throughflow and storage at time $t$, respectively.

The system efficiencies and stress have the potential to play the role in ecological systems of heart rate graphs in examining the human body, as they can detect system disturbances and abnormalities. Rapid unusual fluctuations in the graphs of these functions indicate an excess amount of input into the system, as presented componentwise for certain intercompartmental flows in Case studies~\ref{ex:hallam} and~\ref{ex:disc}. Consequently, the system efficiencies and stress can dynamically quantify the \textit{system resilience} (restoration time) and \textit{resistance}. The maximum \textit{period of the fluctuations} in the \texttt{diact} system efficiencies can be used as a measure for the system resilience. The maximum \textit{amplitude} of the system stress can then be used as a measure for the system resistance--the smaller this amplitude, the more resistant to the impulses the system. Note that the system resilience has unit of time $[\rm{t}]$, and the system resistance is dimensionless. These measures can be used as ecological indicators to monitor ecosystems for environmental impacts and, therefore, may prove useful for environmental assessment and management.

The local and average, flow- and storage-based, {\em simple} and {\em composite} $\texttt{diac}$ {\em effect measures}, {\em indices}, and {\em system efficiencies} for all $\texttt{diac}$ interaction types can be formulated similar to their transfer counterparts, by substituting the corresponding $\texttt{diac}$ flows and storages for the transfer flows and storages in all equations above. We use a tilde notation over the simple versions of the vector and matrix quantities. As examples, the flow-based {\em simple cycling} and {\em composite indirect effect indices} induced by environmental inputs can be written as follows: 
\begin{equation}
\label{eq:diact_eff}
\begin{aligned}
\tilde{\texttt{c}}^\tau(t) & 
= \frac{ \mathbf{1}^T \, \tilde{T}^{\texttt{c}} (t) \,  \bm{1} } { \check{\sigma}^{\tau}(t) } 
\quad \mbox{and} \quad
\texttt{i}^{\tau}(t) 
= \frac{ \mathbf{1}^T \, {T}^{\texttt{i}} (t) \,  \bm{1} } { \check{\sigma}^{\tau}(t) }  .
\end{aligned} 
\end{equation}

The simple and composite \texttt{diact} effect indices can ecologically be interpreted as the direct, indirect, non-cycling, cycling, and total influence of one system compartment on another, induced by a single and all environmental inputs, respectively. More specifically, the flow-based composite transfer effect index, $\texttt{t}^{\tau}_{ik}(t)$, for example, can be interpreted as the total influence of compartment $k$ on $i$ at time $t$, induced by {\em all} environmental inputs during $[t_1,t]$. The flow-based simple transfer effect index, $\texttt{t}^{\tau}_{i_k}(t)=\texttt{t}^{\tau}_{i_k k_k}(t)$, can then be interpreted as the total influence of compartment $k$ on $i$ at time $t$, induced by the {\em single} environmental input, $z_k(t)$, during $[t_1,t]$. Similarly, the flow-based composite transfer effect index, $\texttt{t}^{\tau}_{i_0 k_0}(t)$, can be interpreted as the total influence of compartment $k$ on $i$ at time $t$, induced by {\em all} initial stocks during $[t_1,t]$. All the other flow- and storage-based, simple and composite \texttt{diact} effect indices can be interpreted similarly.

Ecologically, the flow-based composite transfer effect vector $\check{\texttt{t}}^{\tau}(t)$ can be interpreted as the transfer effect of the system on the compartments and $\hat{\texttt{t}}^{\tau}(t)$ as those of the compartments on the system induced by environmental inputs at time $t$. The flow-based simple transfer effect vector $\check{\tilde{\texttt{t}}}^{\tau}(t)$ can then be interpreted as the transfer effect of the system on the compartments and $\hat{\tilde{\texttt{t}}}^{\tau}(t)$ as that of the compartments on the subsystems induced by single environmental inputs at time $t$. All the other \texttt{diac} effect vectors can be formulated as their transfer counterparts given in Eq.~\ref{eq:b_utilityU3} by the corresponding substitutions and interpreted accordingly. The scalar \texttt{diac} system effect indices can also be formulated as their transfer counterparts given in Eq.~\ref{eq:b_effT} by the corresponding substitutions. They can be interpreted as the \texttt{diac} effects of the system on itself induced by environmental inputs.

In static ecological network analyses, Finn's Cycling Index (FCI) is the standard flow-based measure that quantifies cycling system flows \cite{Finn1980}. A storage-based cycling index (SCI) is also formulated in the literature \cite{Ma2014}. A dynamic measure for flow or storage cycling has not been formulated yet. The proposed methodology explicitly formulates the dynamic local and average, simple and composite, flow- and storage-based cycling indices, as well as the corresponding system efficiencies at both compartmental and subcompartmental levels for the first time. It is also shown by \cite{Coskun2017SCSA} that, at steady-state, the proposed dynamic flow- and storage-based simple cycling effect indices at the compartmental level are equivalent to the FCI and SCI, respectively. Static cycling index is sometimes associated with ecosystem stress \cite{Wulff1989}. The cycling flow efficiency will alternatively be called {\em system stress}, accordingly.

In the input-output economics and environ theory, the indirect effects are considered to be flow contributions carried by subsequent steps after the first entrance into a compartment. Even the direct transactions, after the first step, are considered as indirect contribution in various formulations proposed in the literature \cite{Leontief1966,Patten1985a,Patten1985,Ulanowicz1990b,Higashi1989,Borrett2011,Ma2013}. They are, therefore, microscopic quantities that cannot quantify indirect interactions accurately \cite{Coskun2017SESM,Coskun2019ITR}. These static indirect effect indices are formulated in reference to the environmental inputs and without actually defining the indirect flow between any two system compartments. The proposed dynamic indirect effect indices of a compartment on any other are based on the indirect flows and storages introduced through the system decomposition theory with a different derivation rationale than the current static indices. The static versions of the proposed dynamic indices capture experimental system behavior more accurately than the current static formulations, as shown by~\cite{Coskun2017SESM,Coskun2019ITR}.

\subsection{The \texttt{diact} utility measures and indices}
\label{sec:utility}

The $\texttt{diact}$ {\em utility measures and indices} and the corresponding {\em efficiencies} are systematically introduced in this section. In general terms, the dynamic $\texttt{diact}$ utility measures will be defined as the {\em relative} $\texttt{diact}$ effects of one compartment on another. The subcompartmental level formulations in parallel are straightforward, by using subflows and associated substorages instead of flows and storages. 

We will first define the dynamic {\em transfer utility measures and indices}. The {\em flow-} and {\em storage-based composite transfer utility indices} of a set of compartments $K$ to another set $I$, $\mathbbm{t}^\tau_{IK}(t)$ and $\mathbbm{t}^x_{IK}(t)$, quantify the relative flow- and storage-based transfer effects of $K$ on $I$ induced by environmental inputs at time $t$. They measure the normalized relative net benefit ($\mathbbm{t}^\tau_{IK}(t) >0$ and $\mathbbm{t}^x_{IK}(t) >0$) or harm ($\mathbbm{t}^\tau_{IK}(t) <0$ and $\mathbbm{t}^x_{IK}(t) <0$), that is transmitted from the set of compartments $K$ to $I$ at time $t$ based on their respective net gains (inflows and associated storages) or losses (outflows and associated storages). The composite transfer utility indices induced by environmental inputs will be formulated as
\begin{equation}
\label{eq:b_effU}
\begin{aligned}
{\mathbbm{t}}^\tau_{IK}(t) = 
\texttt{t}^\tau_{IK}(t) - \texttt{t}^\tau_{KI}(t) 
\quad \mbox{and} \quad
{\mathbbm{t}}^x_{IK}(t) = 
\texttt{t}^x_{IK}(t) - \texttt{t}^x_{KI}(t) 
\end{aligned}
\end{equation}
where $I,K \subseteq \{ 1,\ldots,n \}$. If the sets $I$ and $K$ have one element, that is, $I=\{ i\}$ and $K=\{ k\}$, these indices will be denoted by $\mathbbm{t}^\tau_{ik}(t)$ and $\mathbbm{t}^x_{ik}(t)$. Note that since the transfer effect indices are dimensionless, the transfer utility indices are also dimensionless. The flow- and storage-based composite transfer utility indices, $\mathbbm{t}_{i_0 k_0}^\tau(t)$ and $\mathbbm{t}_{i_0 k_0}^x(t)$, of initial subcompartment $k_0$ to $i_0$ induced by the initial stocks can be formulated similarly by using  the corresponding composite transfer effect indices, $\texttt{t}_{i_0 k_0}^\tau(t)$ and $\texttt{t}_{i_0 k_0}^x(t)$, respectively.

The {\em flow-} and {\em storage-based composite transfer utility matrix measures} induced by environmental inputs are denoted by $\mathbbm{T}^\tau(t) = \left( \mathbbm{t}^\tau_{ik}(t) \right)$ and $\mathbbm{T}^x(t) = \left( \mathbbm{t}^x_{ik}(t) \right)$ and formulated in matrix form as
\begin{equation}
\label{eq:b_effU2}
\begin{aligned}
\mathbbm{T}^\tau(t) = \frac{ 1} { \check{\sigma}^\tau(t) } \left( {T}^{\texttt{t}}(t) - {{T}^{\texttt{t}}}^T(t) \right) \quad \mbox{and} \quad 
\mathbbm{T}^x(t) = \frac{ 1} { \sigma^x(t) } \left( {X}^{\texttt{t}}(t) - {{X}^{\texttt{t}}}^T(t) \right) .
\end{aligned}
\end{equation}
The {\em flow-based composite transfer utility} of the system to the compartments, $\check{\mathbbm{t}}^\tau(t)$, and that of the compartments to the system, $\hat{\mathbbm{t}}^\tau(t)$, induced by environmental inputs will be defined as {\em vector measures}:
\begin{equation}
\label{eq:b_effU3}
\begin{aligned}
\check{\mathbbm{t}}^\tau(t) = \mathbbm{T}^\tau(t) \, \mathbf{1} \quad \mbox{and} \quad \hat{\mathbbm{t}}^\tau(t) = \mathbf{1}^T \, \mathbbm{T}^\tau(t) \quad \mbox{with} \quad 
\check{\mathbbm{t}}^\tau(t) = - \left( \hat{\mathbbm{t}}^\tau(t) \right)^T .
\end{aligned}
\end{equation}
The last relationship in Eq.~\ref{eq:b_effU3} is due to the fact that $\mathbbm{T}^\tau(t)$ and $\mathbbm{T}^x(t)$ are skew-symmetric matrices, i.e., ${\mathbbm{T}^\tau}^T(t) = -\mathbbm{T}^\tau(t)$ and ${\mathbbm{T}^x}^T(t) = -\mathbbm{T}^x(t)$. The {\em storage-based composite transfer utility vectors} can be defined similarly. Due to the skew-symmetry, the {\em flow-} and {\em storage-based composite transfer utility indices} for the system induced by environmental inputs are zero: $$\mathbbm{t}^\tau(t) = \mathbf{1} \, \mathbbm{T}^\tau(t) \, \mathbf{1} = 0 \quad \mbox{and} \quad \mathbbm{t}^x(t) = \mathbf{1} \, \mathbbm{T}^x(t) \, \mathbf{1} = 0.$$
These relationships are true for all $\texttt{diact}$ {\em utility matrix measures}. 

The {\em average composite transfer utility indices} induced by environmental inputs, which can be formulated similar to the average effect indices defined in Section~\ref{sec:effects}, are also zero, due to the skew-symmetry of the corresponding matrix measures. That is,
\begin{equation}
\label{eq:cindex_intT_ave}
\mathbbm{t}^\tau(t_1,t) = 0
\quad \mbox{and} \quad 
\mathbbm{t}^x(t_1,t) = 0 . 
\end{equation}
The {\em flow-} and {\em storage-based composite transfer utility efficiencies} induced by environmental inputs are then defined as the time derivatives of the corresponding utility indices as follows:
\begin{equation}
\label{eq:b_effU4}
\begin{aligned}
\dot{\mathbbm{t}}^\tau_{ik}(t) = 
\dot{\texttt{t}}^\tau_{ik}(t) - \dot{\texttt{t}}^\tau_{ki}(t) 
\quad \mbox{and} \quad
\dot{\mathbbm{t}}^x_{ik}(t) = 
\dot{\texttt{t}}^x_{ik}(t) - \dot{\texttt{t}}^x_{ki}(t) .
\end{aligned}
\end{equation}

The flow- and storage-based, {\em simple} and {\em composite} $\texttt{diact}$ {\em utility measures}, {\em indices}, and {\em system efficiencies} for all $\texttt{diact}$ interaction types can be formulated, similar to their transfer counterparts, by substituting the corresponding $\texttt{diac}$ flows and storages for the transfer flows and storages in all equations above. We use a tilde notation over the simple versions of the vector and matrix quantities. 

The simple and composite \texttt{diact} utility indices can ecologically be interpreted as the relative direct, indirect, non-cycling, cycling, and total influence of system compartments on each other, induced by single and all environmental inputs, respectively. More specifically, the flow-based composite transfer utility index, $\mathbbm{t}^{\tau}_{ik}(t)$, for example, can be interpreted as the relative total influence of compartment $k$ on $i$ at time $t$, induced by {\em all} environmental inputs during $[t_1,t]$. The flow-based simple transfer utility index, $\mathbbm{t}^{\tau}_{i_k}(t) = \mathbbm{t}^{\tau}_{i_k k_k}(t)$, can then be interpreted as the relative total influence of compartment $k$ on $i$ at time $t$, induced only by the {\em two} corresponding environmental inputs, $z_k(t)$ and $z_i(t)$, during $[t_1,t]$. Similarly, the flow-based composite transfer utility index, $\mathbbm{t}^{\tau}_{i_0 k_0}(t)$, can be interpreted as the relative total influence of compartment $k$ on $i$ at time $t$, induced by {\em all} initial stocks during $[t_1,t]$. All the other flow- and storage-based, simple and composite \texttt{diact} utility indices can be interpreted similarly.

Ecologically, the flow-based composite transfer utility vector $\check{\mathbbm{t}}^{\tau}(t)$ can be interpreted as the relative transfer effects of a system on its compartments and $\hat{\mathbbm{t}}^{\tau}(t)$ as those of the compartments on the system induced by environmental inputs at time $t$. The flow-based simple transfer utility vectors can be interpreted similarly. All the other \texttt{diac} utility vectors can also be formulated as their transfer counterparts given in Eq.~\ref{eq:b_effU3} by the corresponding substitutions and interpreted accordingly.

A direct utility index was introduced in the literature for static systems \cite{Patten1991,Fath1999} motivated by a methodology introduced by \cite{Ulanowicz1990b}. The local, compartmental normalization in this formulation makes the physical interpretation of the utility index difficult as a system measure. The static version of the proposed direct utility index is compared with this utility index by \cite{Coskun2017SESM}. The proposed index is different from the authors' static formulation, due to its global normalization procedure in accordance with the $\texttt{diact}$ effect index formulations introduced in Section~\ref{sec:effects}. This global normalization allows for local interpretations of intercompartmental dynamics relative to the entire system.

\subsection{The \texttt{diact} exposures and residence times}
\label{sec:EEtime}

The impact of environment on system compartments can be evaluated by their exposure to environmental inputs. The exposure to ionizing radiation, poisons, and other bioactive chemical agents are important topics of concern for human health and welfare. In this section, we introduce the dynamic {\em exposure} and {\em residence time} measures and indices. These novel mathematical system analysis tools can find use in radiobiology, toxicology, pharmacokinetics, and other applied environmental and medical fields.

The {\em exposure} of compartment $i$ during $[t_1,t]$, $t_1 \geq t_0$, to the environmental input into component $k$, $z_k(t)$, can be defined component-wise as
\begin{equation}
\label{eq:exposure}
e_{i_k}(t_1,t) = \int_{t_1}^{t} x_{i_k}(s) \, ds
\end{equation}
for $i=1,\ldots,n$, and $k=0,\ldots,n$. Note that, the unit of exposure is mass $\times$ time, $[\rm{m \, t}]$. The unit of storage can be replaced by the unit of the conserved quantity in question, such as energy or currency, depending on the model of interest. Excluding exposure of the initial subcompartments within the initial subsystem ($k=0$) to the system flows derived from the initial stocks, the $n \times n$ {\em exposure matrix}, $E_{ik}(t_1,t)=\left( e_{i_k}(t_1,t) \right)$, can be expressed in matrix form as
\begin{equation}
\label{eq:exposure2}
E(t_1,t) = \int_{t_1}^{t} X(s) \, ds .
\end{equation}
The {\em exposure} of compartments, $\check{e}(t_1,t)$, and subsystems, $\hat{e}(t_1,t)$, to the environmental inputs during $[t_1,t]$ can then be formulated as vector functions:
\[ \check{e}(t_1,t) = E(t_1,t) \, \mathbf{1} \quad \mbox{and} \quad \hat{e}(t_1,t) = \mathbf{1}^T \, E(t_1,t) . \]
The scalar {\em system exposure index} to environmental inputs during $[t_1,t]$ can also be formulated as
\begin{equation}
\label{eq:exposure3}
e(t_1,t) = \mathbf{1}^T E(t_1,t) \, \mathbf{1}.
\end{equation}

We also define the {\em exposure time} or {\em residence time} of the storage in compartment $i$ at time $t$ as 
\begin{equation}
\label{eq:irestime}
\begin{aligned}
{r}_{i}(t,x) = \frac{x_{i}(t)}{\hat{\tau}_{i}(t,{\rm x})} = \frac{x_{i_i}(t)}{ \hat{\tau}_{i_i}(t,{\rm x})} = \frac{x_{i_k}(t)}{ \hat{\tau}_{i_k}(t,{\rm x})}
\end{aligned}
\end{equation} 
for $i=1,\ldots,n$ and $k=0,\ldots,n$, where the denominators are nonzero. The residence time of substorage in subcompartment $i_k$ at time $t$ are the same for any $k$, as indicated in Eq.~\ref{eq:irestime}. Therefore, excluding the initial subsystem, the $n \times n$ diagonal matrix function
$$\mathcal{R}(t,x) = \diag{([r_{1} (t,x),\ldots,r_{n} (t,x)])} $$ 
will be called the {\em residence time matrix}. It can then be expressed in the following various forms:
\begin{equation}
\label{eq:irestimeM}
\begin{aligned}
\mathcal{R}(t,x)  = \mathcal{X}(t) \, \mathcal{T}^{-1}(t,{x}) 
= \mathcal{X}_k(t,{\rm x}) \, \mathcal{\hat T}_k^{-1}(t,{\rm x})  = X(t) \, \hat{T}^{-1}(t,{\rm x}) 
\end{aligned}
\end{equation} 
as formulated in Eq.~\ref{eq:matrix_A} \cite{Coskun2017DCSAM,Coskun2017NDP}. The diagonal $k^{th}$ {\em substorage}, {\em inward}, and {\em outward subthroughflow matrices} used in Eq.~\ref{eq:irestimeM} are defined as
\begin{equation}
\label{eq:res_comps}
\begin{aligned} 
\mathcal{X}_k(t) & = \diag \left( [ x_{1_k}(t), \ldots, x_{n_k}(t) ] \right), \quad \mbox{and} \quad \\
\check{\mathcal{T}}_k(t,{\rm x}) & = \diag{\left([\check{\tau}_{1_k}(t,{\rm x}),\ldots,\check{\tau}_{n_k}(t,{\rm x})] \right)} , \quad 
\hat{\mathcal{T}}_k(t,{\rm x}) = \diag{ \left( [\hat{\tau}_{1_k}(t,{\rm x}),\ldots,\hat{\tau}_{n_k}(t,{\rm x})] \right) } ,
\end{aligned}
\nonumber
\end{equation} 
for the $k^{th}$ subsystem, $k=0,\ldots,n$.

The $i^{th}$ diagonal entry of $\mathcal{R}(t,{x})$ at time $t_1$, ${r}_{i} (t_1,{x})$, can be interpreted as the time required for the outward throughflow, at the constant rate of $\hat{\tau}_i(t_1,{x})$, to completely empty compartment $i$ with the storage of $x_{i}(t_1)$. The diagonal structure of the residence time matrix indicates that all subcompartments of compartment $i$ vanish simultaneously.

Ecologically, $\mathcal{R}(t,x)$ can be used as a quantitative ecosystem indicator that represents the {\em compartmental activity levels}: the smaller the residence time, the more active the corresponding compartment. The derivative of the residence time matrix, $\dot{\mathcal{R}}(t,x)$, will be called the {\em reverse activity rate} matrix. Note that the unit of the residence time is time, $[\rm{t}]$, and its time derivative is dimensionless.

The exposure of compartments to the transient and \texttt{diact} flows can be formulated by substituting the corresponding transient and \texttt{diact} storages for substorage, $x_{i_k}(t)$, in Eq.~\ref{eq:exposure}. The exposure of compartments to the transient and \texttt{diact} flows will be called the {\em transient} and \texttt{diact} {\em exposures} and be denoted by superscripts $w$ and {\em \texttt{diact}} symbols, respectively. For a given subflow path $p^w_{n_k i_k}= i_k \mapsto j_k \to \ell_k \to n_k$, the transient exposure of subcompartment $\ell_k$ at time $t$ to transient inflow $f^w_{\ell_k j_k i_k}$ along path $p^w_{n_k i_k}$ during $[t_1,t]$, for example, can be formulated as
\begin{equation}
\label{eq:exposure_apx}
e^w_{\ell_k}(t_1,t) = \int_{t_1}^{t} x^w_{n_k \ell_k j_k}(s) \, ds .
\end{equation}
Note that the residence times of the transient and \texttt{diact} storages in subcompartment $\ell_k$, ${r}^{w}_{\ell_k} (t)$ and ${r}^{*}_{\ell_k} (t)$, are equal to $r_{\ell}(t,x)$, due to the equivalence of the outward throughflow and subthroughflow intensities. The {\em transient residence time}, for example, is 
$$ {r}_{\ell} (t,x) = {r}^{w}_{\ell} (t) = { x^w_{n_k \ell_k j_k}(t) }/{\hat{\tau}^w_{\ell_k}(t)} $$
where $\hat{\tau}^w_{\ell_k}(t)$ is the cumulative transient subflow~\cite{Coskun2017NDP}.
   
It is also worth nothing that the \texttt{diact} {\em exposures} can be interpreted as unnormalized, storage-based, average \texttt{diact} effects. The {\em indirect exposure} of compartment $i$ at time $t$ to the composite flow transmitted from $k$ indirectly through other compartments during $[t_1,t]$, for example, can be formulated as
\begin{equation}
\label{eq:diact_exposure}
\emph{\texttt{i}}_{i k}(t_1,t) = \int_{t_1}^{t} \, x^\texttt{i}_{ik} (s) \, ds =
\texttt{i}_{i k}^x(t_1,t) \, { \int_{t_1}^{t} \sigma^{x}(s) \, ds } .
\end{equation}

Illustrative examples for the system measures and indices introduced in this section are presented in Section~\ref{sec:results}. 

\subsection{Quantitative definitions of interspecific interactions}
\label{sec:qdii}

One of the immediate potential ecological applications of the system decomposition theory is the quantitative analysis of food webs. Community ecology classifies interspecific interactions qualitatively by the network topology without regard for system flows \cite{Menge1995,Menge1997}. Increasing complexity of intricate food webs in most cases disallows this structural determination due to various factors, such as multiple food chains of potentially different lengths between two species \cite{Holt1997,Wootton1994,Patten2007}.

The signs of the \texttt{diact} flows, storages, or effect indices \textit{graph theoretically} indicate the existence of the directed \texttt{diact} \textit{paths} between the corresponding nodes or vertices (compartments). For example, the relationship $\sgn{(\tau^\texttt{c}_{i_k}(t))} > 0$ indicates that there is at least one closed path from node $i$ to itself at time $t$, where $\sgn(\cdot)$ is the sign function. Similarly, the relationship $\sgn{(\tau^\texttt{i}_{ij}(t))} > 0$ shows that there is at least one indirect path from node $j$ through other nodes to $i$ at time $t$. In the context of interspecific interactions, this implies that there is a food chain from species $j$ indirectly to $i$ at time $t$.

A mathematical technique for the {\em sign} and {\em strength} analysis of the \texttt{diact} {\em interactions} for dynamic systems modeling food webs has been developed recently by \cite{Coskun2017DCSAM}. The {\em sign} and {\em strength} of the \texttt{diact} interactions induced by environmental inputs between species $i$ and $j$ were defined respectively as follows:
\begin{equation}
\label{eq:iidm}
\delta^\texttt{*}_{ij}(t) = \sgn{(\tau^\texttt{*}_{ij}(t) - \tau^\texttt{*}_{ji}(t) )} \quad \mbox{and} \quad \mu^\texttt{*}_{ij}(t) = 
\frac{| \tau^\texttt{*}_{ij}(t) - \tau^\texttt{*}_{ji}(t) | } { \check{\tau}_{i}(t,x) + \check{\tau}_{j}(t,x) } 
\end{equation} 
where the superscript ($^\texttt{*}$) represents any of the \texttt{diact} symbols. Following the convention of community ecology, instead of $(+1)$ and $(-1)$, $(+)$ and $(-)$ notations will be used for the sign of the \texttt{diact} interactions. The strength, $0 \leq \mu^\texttt{*}_{ij}(t) \leq 1$, is defined to be zero, if both terms in its denominator are zero. 

For the analysis of \texttt{diact} interactions ranging from the individual and local to the system-wide and global scale, the strength of the interactions can be formulated with the normalization by $\tau^\texttt{*}_{ij}(t) + \tau^\texttt{*}_{ji}(t)$, $\tau^\texttt{t}_{ij}(t) + \tau^\texttt{t}_{ji}(t)$, $\check{\tau}_{i}(t,x) + \check{\tau}_{j}(t,x)$, as in Eq.~\ref{eq:iidm}, or $\check{\sigma}^{\tau}(t) = \bm{1}^T \, \check{\tau}(t,x)$ in the given order. At the global scale, the sign and strength of the local direct interactions between species $i$ and $j$ induced by environmental inputs, for example, can accordingly be formulated as
\begin{equation}
\label{eq:iidm2}
\delta^\texttt{d}_{ij}(t) = \sgn{( \mathbbm{d}^\tau_{ij}(t) )} 
\quad \mbox{and} \quad 
\mu^\texttt{d}_{ij}(t) = \frac{| \tau^\texttt{d}_{ij}(t) - \tau^\texttt{d}_{ji}(t) | } { \check{\sigma}^{\tau}(t) } = | \mathbbm{d}^\tau_{ij}(t) | 
\end{equation} 
using the utility indices. The direct {\em neutral relationship} between species $i$ and $j$ and ``{\em predation}'' of species $i$ on $j$ can quantitatively be characterized, respectively, as follows:
\begin{equation}
\label{eq:ii}
\mathbbm{d}_{ij}^\tau(t) = 0 \, \, \, \Rightarrow \, \, \, \delta^\texttt{d}_{ij}(t) = (0) 
\quad \mbox{and} \quad 
\mathbbm{d}_{ij}^\tau(t) > 0 \, \, \, \Rightarrow \, \, \, \delta^\texttt{d}_{ij}(t) = (+) .
\end{equation} 
The sign and strength of the other \texttt{diact} interactions, as well as their characterization can be formulated similarly by using the corresponding \texttt{diact} flows in Eqs.~\ref{eq:iidm} and~\ref{eq:ii} instead of the direct flows.
\begin{table}
     \centering
     \caption{Quantitative definitions of interspecific interactions.}     
     \label{tab:qdii}
\resizebox{\linewidth}{!}{         
     \begin{tabular*}{\textwidth}{p{.18\textwidth} | p{.09\textwidth} | p{.13\textwidth} | p{.14\textwidth} | p{.3\textwidth}}  
\hline      
\multicolumn{1}{ c | }{Type} & \multicolumn{3}{  c }{Definition} &\multicolumn{1}{ | c } {Strength} \\ 
\hline 
     \noalign{\vskip 2pt} 
Neutralism & 
$\mathbbm{d}^\tau_{ij} = 0 $
& $\texttt{d}^\tau_{ik} \, \texttt{d}^\tau_{jk} = 0$ 
& $\mathbbm{i}^\tau_{ij} = 0$  
& $\mu_{ij}^n = 0$  \\ 
Mutualism & 
$\mathbbm{d}^\tau_{ij} = 0$ & 
$\texttt{d}^\tau_{ik} \, \texttt{d}^\tau_{jk} = 0$ & 
$\mathbbm{i}^\tau_{ij} \neq 0$ & 
$ \mu_{ij}^m = ( \tau^\texttt{i}_{ij} + \tau^\texttt{i}_{ji}  ) / ( \check{\tau}_{i} + \check{\tau}_{j}  )  $ \\
Commensalism & 
$\mathbbm{d}^\tau_{ij} = 0 $ &
$\texttt{d}^\tau_{ik} \, \texttt{d}^\tau_{jk} \neq 0 $ &
$\mu_{ij,k}^c \gg 1/2 $ 
& \multirow{2}{*} {\noindent $ \mu_{ij,k}^c = | \tau^\texttt{d}_{ik} - \tau^\texttt{d}_{jk} | / ( \tau^\texttt{d}_{ik} + \tau^\texttt{d}_{jk} ) $ } \\
Competition & 
$\mathbbm{d}^\tau_{ij} = 0 $ &
$\texttt{d}^\tau_{ik} \, \texttt{d}^\tau_{jk} \neq 0 $ &
$\mu_{ij,k}^c \ll 1/2 $ & 
\\
Exploitation  &  
$\texttt{d}^\tau_{ij} > 0 $ &
$ \texttt{d}^\tau_{ji} = 0$ & & 
$  \mu_{ij}^e = \tau^\texttt{d}_{ij} / \hat{\tau}_{j} $ \\
   \noalign{\vskip 2pt} 
\hline 
     \end{tabular*}
     }
\end{table}

A mathematical technique for the characterization and classification of the main interspecific interaction types, such as neutralism, mutualism, commensalism, competition, and exploitation in static food webs has also recently been developed based on the \texttt{diact} flows and storages \cite{Coskun2017SESM}. Following the same rationale, the quantitative definitions of the main interspecific interactions induced by environmental inputs and their strength are extended to nonlinear dynamic food webs in Table~\ref{tab:qdii}. The function arguments $x$ and $t$ are dropped in these formulations in the table for readability. The strength of mutualism and exploitation can be reformulated using the effect indices for the analysis of interspecific interactions at the global scale as follows:
\begin{equation}
\label{eq:iidm3}
\begin{aligned}
\mu^{m}_{ij}(t) & = \frac{ \tau^\texttt{i}_{ij}(t) + \tau^\texttt{i}_{ji}(t) }{ \check{\sigma}^{\tau}(t) } = \texttt{i}^\tau_{ij}(t) + \texttt{i}^\tau_{ji}(t)
\quad \mbox{and} \quad \\
\mu^{e}_{ij}(t) & = \frac{ \tau^\texttt{d}_{ij}(t) }{ \hat{\sigma}^{\tau}(t) } 
\quad \mbox{or} \quad 
\mu^{e}_{ij}(t) = \frac{ \tau^\texttt{d}_{ij}(t) }{ \check{\sigma}^{\tau}(t) } = \texttt{d}_{ij}(t)  .
\end{aligned}
\end{equation}

The classification of both the \texttt{diact} and main interspecific interactions induced by the initial stocks can be formulated by using the composite \texttt{diact} subflows for the initial subsystem, $\tau^\texttt{*}_{i_0j_0}(t)$, instead of $\tau^\texttt{*}_{ij}(t)$ in the corresponding formulations. The storage-based quantitative definition of the \texttt{diact} and main interspecific interactions can also be formulated in parallel by substituting the \texttt{diact} storages for the corresponding \texttt{diact} flows. We will use superscript $x$ to distinguish the storage-based sign and strength measures. The storage-based formulations represent the history of interspecific interactions during $[t_1,t]$, while the flow-based formulations represent simultaneous interactions at time $t$. Lastly, for the classification of the interspecific interactions induced by individual environmental inputs, the simple \texttt{diact} flows and storages can be used instead of their composite counterparts. A tilde notation will be used over the simple versions of the sign and strength measures.

\section{Results}
\label{sec:results}

The proposed dynamic methodology is applied to various discrete and continuous ecological models from the literature. The dynamic measures and indices formulated above for the \texttt{diact} effects, utilities, exposures, and residence times, as well as the corresponding system efficiencies, stress, resilience, and resistance together with their ecological implications are presented for these models in this section.

The results indicate that the proposed methodology precisely quantifies dynamic system functions, properties, and behaviors, effectively determine the environmental influence on system compartments and intercompartmental dynamics, is sensitive to perturbations due to even a brief unit impulse, and, thus, can be used for rigorous dynamic analysis of nonlinear ecological systems. It is worth noting, however, that this present work proposes a mathematical {\em method}\textemdash a systematic technique designed for analyzing dynamic nonlinear ecosystem models using the proposed measures and indices as ecosystem indicators\textemdash and it is not itself a {\em model}. Therefore, we focus more on demonstrating the efficiency and wide applicability of the mathematical system analysis tools introduced as quantitative ecological indicators in this paper. It is expected that once the method is accessible to a broader community of environmental ecologists, it can be used for the holistic analysis of specific models of interest.

\subsection{Case study}
\label{ex:hallam}

A nonlinear model introduced by \cite{Hallam1985} was recently analyzed through the system decomposition~\cite{Coskun2017DCSAM}. In particular, the substorages and subthroughflows, as well as the transient and \texttt{diact} flows and storages were presented in that article. In this case study, the dynamic measures and indices introduced in the present manuscript are provided for this ecosystem model together with their ecological interpretations.
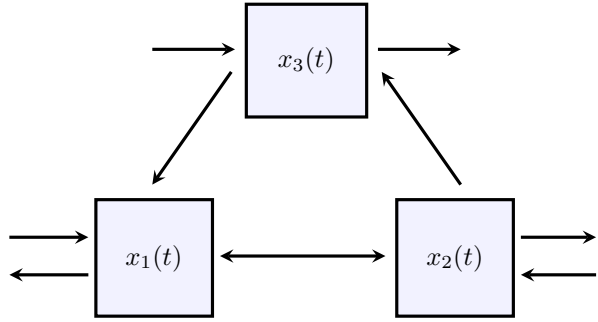
\begin{figure}[t]
\begin{center}
\begin{tikzpicture}
\centering
   \draw[very thick,  fill=blue!5, draw=black] (-.05,-.05) rectangle node(R1) {$x_1(t)$} (1.5,1.5) ;
   \draw[very thick,  fill=blue!5, draw=black] (3.95,-.05) rectangle node(R2) {$x_2(t)$} (5.5,1.5) ;   
   \draw[very thick,  fill=blue!5, draw=black] (1.95,2.6) rectangle node(R3) {$x_3(t)$} (3.55,4.1) ;         
       \draw[very thick,stealth-stealth,draw=black]  (1.6,.75) -- (3.8,.75) ;  
       \draw[very thick,stealth-,draw=black] (0.7,1.7) -- (1.75,3.2) ;     
       \draw[very thick,-stealth,draw=black]  (4.8,1.7) -- (3.75,3.2) ;  
       \draw[very thick,stealth-,draw=black]  (5.6,.5) -- (6.6,.5) ;      
       \draw[very thick,-stealth,draw=black]  (5.6,1) -- (6.6,1) ;         
       \draw[very thick,stealth-,draw=black]  (-1.2,.5) -- (-0.15,.5) ;  
       \draw[very thick,-stealth,draw=black]  (-1.2,1) -- (-0.15,1) ;    
       \draw[very thick,-stealth,draw=black]  (.7,3.5) -- (1.8,3.5) ;    
       \draw[very thick,-stealth,draw=black]  (3.7,3.5) -- (4.8,3.5) ;   
\end{tikzpicture}
\end{center}
\caption{Schematic representation of the model network (Case study~\ref{ex:hallam}).}
\label{fig:hallam_diag}
\end{figure}

The resource-producer-consumer model by \cite{Hallam1985} consists of the dynamics for three components: $x_1(t) = r(t)$ is the nutrient storage (such as phosphorus or nitrogen) present at time t; $x_2(t) = s(t)$ represents the nutrient storage in the producer (such as phytoplankton) population; and $x_3(t) = c(t)$ denotes the nutrient storage in the consumer (such as zooplankton) population (see Fig.~\ref{fig:hallam_diag}). The conservation of nutrient is the basic model assumption. The system flows are described as follows:
\begin{equation}
\label{eq:hallam_flows}
\begin{aligned} 
F(t,x) = 
\begin{bmatrix}
    0  & d_1 \, s(t) & d_2 \, c(t)  \\ \displaystyle
    \frac{\alpha_1 \, s(t) \, r(t) }{\alpha_2 + r(t) }  & 0 & 0 \\ 
    0 & \displaystyle \frac{\beta_1 \, s(t) \, c(t) }{\beta_2 + s(t) } & 0 
\end{bmatrix},
\, \, \,
z(t) = 
\begin{bmatrix}
    z_1(t) \\
    z_2(t) \\
    z_3(t) 
\end{bmatrix},
\, \, \,
y(t) = 
\begin{bmatrix}
    r(t) \\
    s(t) \\
    c(t) 
\end{bmatrix}
\end{aligned}
\nonumber
\end{equation}
where the parameters are given as
\[d_1=2.7, \, d_2=2.025, \, \alpha_2=0.098, \, \beta_1=2, \, \beta_2=20, \, \mbox{and } \, \alpha_1=1.\]
The environmental input is $z(t) = [1, {\mathrm{e}}^{\frac{-(t-15)^2}{2}} +0.1, 1]^T$. That is, while $z_1(t)=1$ and $z_3(t)= 1$ are constant, the system is perturbed with a with time-dependent Gaussian impulse function $z_{2}(t) = {\mathrm{e}}^{\frac{-(t-15)^2}{2}}+0.1$ at the producer compartment. The system of governing equations can be written componentwise as follows:
\begin{equation}
\label{eq:hallam_ex}
\begin{aligned} 
\dot r(t) &= -r(t) + d_1 \, s(t) + d_2 \, c(t) - \frac{\alpha_1 \, s(t) \, r(t)}{\alpha_2 + r(t)} + z_1(t) \\
\dot s(t) &= -(1+d_1) \, s(t) + \frac{\alpha_1 \, s(t) \, r(t)}{\alpha_2 + r(t)} - \frac{\beta_1 \, c(t) \, s(t)}{\beta_2 + s(t)} + z_2(t) \\
\dot c(t) &= -(1+d_2) \, c(t) + \frac{\beta_1 \, c(t) \, s(t)}{\beta_2 + s(t)} + z_3(t)
\end{aligned}
\nonumber
\end{equation}
with the initial conditions of $[r_0,s_0,c_0] = [1, 1, 1]$.
\begin{figure}[t]
    \centering
    \begin{subfigure}[b]{0.46\textwidth}
        \includegraphics[width=\textwidth]{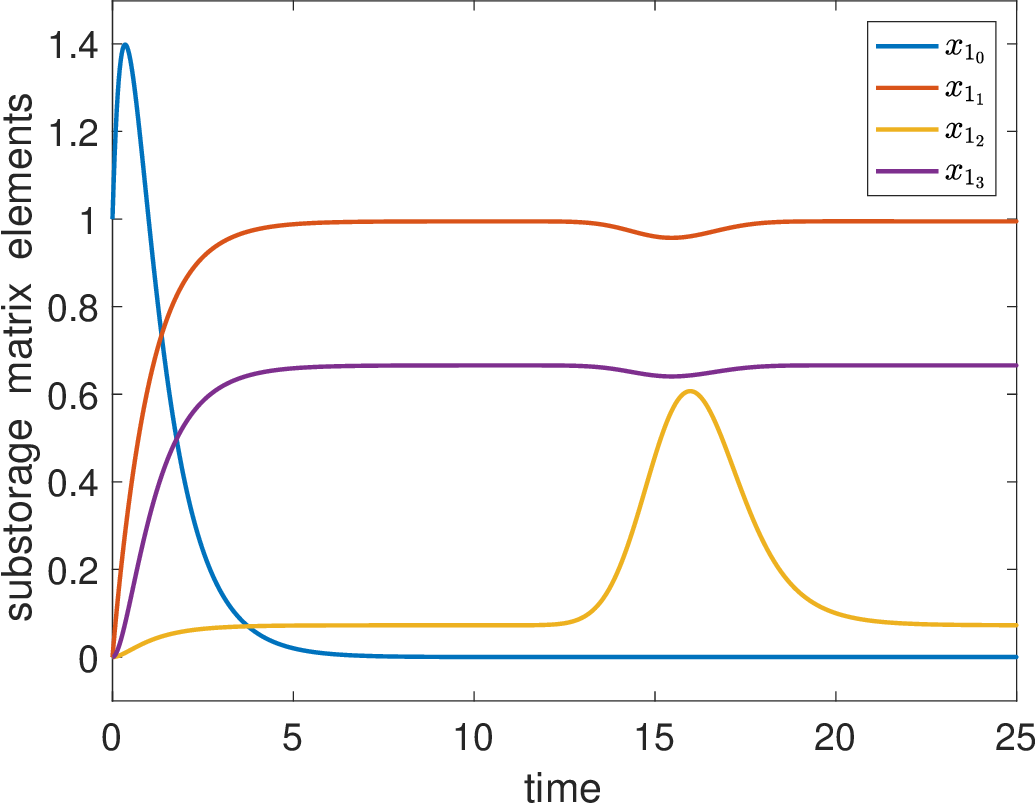}
        \caption{substorages of $x_1(t)$}
        \label{fig:x_1g}
    \end{subfigure}
    \begin{subfigure}[b]{0.45\textwidth}
        \includegraphics[width=\textwidth]{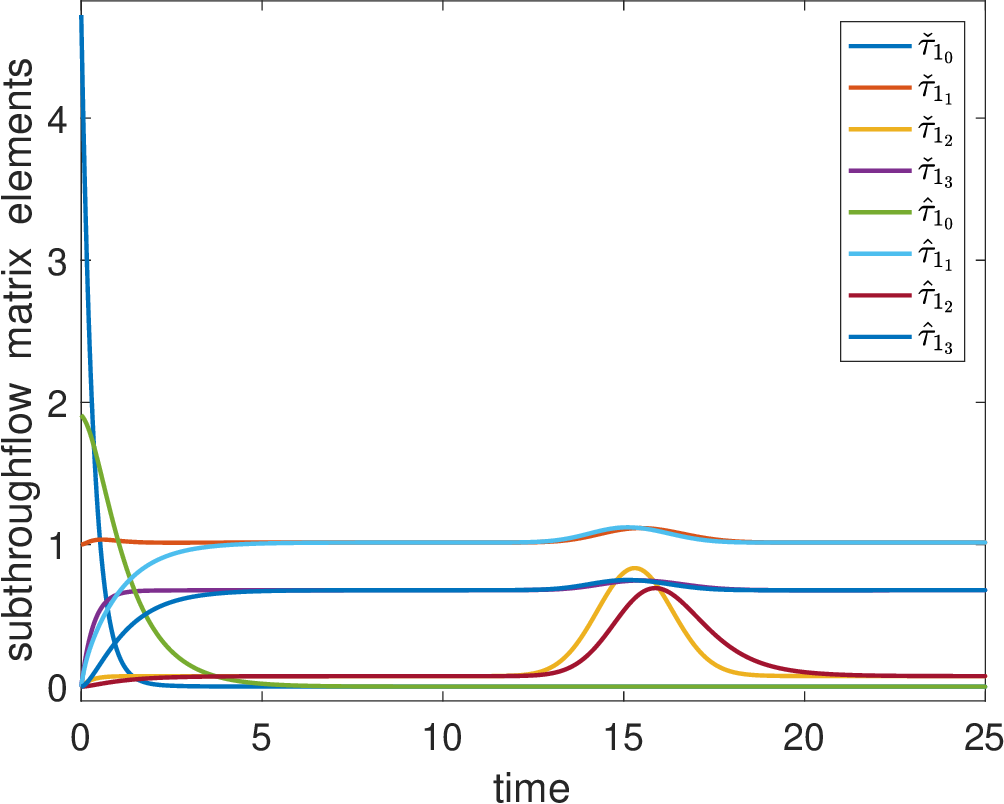}
        \caption{subthroughflows of $\check{\tau}_1(t,{\rm x})$ and $\hat{\tau}_1(t,{\rm x})$}
        \label{fig:tau_1g}
    \end{subfigure} 
        \caption{The numerical results for the selected elements (first rows) of the substorage, $X(t)$, and subthroughflow matrix functions, $\check{T}(t,{\rm x})$ and $\hat{T}(t,{\rm x})$, and the initial substorage, $x_{0}(t)$, and subthroughflow vectors, $\check{\tau}_{0}(t,{\mathrm x})$ and $\hat{\tau}_{0}(t,{\mathrm x})$ (Case study~\ref{ex:hallam}).}
    \label{fig:hallam_g1}
\end{figure}

The subcompartmentalization yields
\[ x_{1_k}(t) = r_k(t), \quad x_{2_k}(t) = s_k(t), \, \, \,  \mbox{and} \, \, \, x_{3_k}(t) = c_k(t) \, \, \,  \mbox{with} \, \, \,  
x_i(t) = \sum_{k=0}^3 x_{i_k} (t) . \]
The flow partitioning then gives the flow regime for each subsystem:
\begin{equation}
\label{eq:hallam_flows_SC}
\begin{aligned} 
F_k(t,{\rm x}) = 
\begin{bmatrix}
    0  & d_{2_k} \, d_1 \, s & d_{3_k} \, d_2 \, c  \\ \displaystyle
    d_{1_k} \, \frac{\alpha_1 \, s \, r}{\alpha_2 + r}  & 0 & 0 \\
    0 &\displaystyle  d_{2_k} \, \frac{\beta_1 \, s \, c}{\beta_2 + s} & 0 
\end{bmatrix},
\, 
\check{z}_k(t,{\rm x}) = 
\begin{bmatrix}
    \delta_{1k} \, z_1 \\
    \delta_{2k} \, z_2 \\
    \delta_{3k} \, z_3
\end{bmatrix},
\, 
\hat{y}_k(t,{\rm x}) = 
\begin{bmatrix}
    d_{1_k} \, r \\
    d_{2_k} \, s \\
    d_{3_k} \, c
\end{bmatrix} 
\end{aligned}
\nonumber
\end{equation}
where $F_k$, $\check{z}_k$, and $\hat{y}_k$ describe the $k^{th}$ direct flow matrix, input, and output vectors for the $k^{th}$ subsystem, and the decomposition factors $d_{i_k}({\rm x})$ are defined by Eq.~\ref{eq:cons2_1new}. Therefore, the dynamic system partitioning methodology yields the following system of governing equations:
\begin{equation}
\label{eq:hallam_exsc}
\begin{aligned} 
{\dot r}_{k}(t) &=  \delta_{1k} \, z_1(t) + d_1 \, s_{k}(t) + d_2 \, c_{k}(t) - r_{k}(t) -  \frac{\alpha_1 \, s(t) \, r_{k}(t) }{\alpha_2 + r(t)}  \\
{\dot s}_{k}(t) &= \delta_{2k} \, z_2(t) + \frac{\alpha_1 \, s(t) \, r_{k}(t)}{\alpha_2 + r(t)} - s_{k}(t) - d_1 \, s_{k}(t) - \frac{\beta_1 \, c(t) \, s_{k}(t) }{\beta_2 + s(t)}  \\
{\dot c}_{k}(t) &= \delta_{3k} \, z_3(t) +  \frac{\beta_1 \, c(t) \, s_{k}(t) }{\beta_2 + s(t)} - c_{k}(t) - d_2 \, c_{k}(t) 
\end{aligned}
\nonumber
\end{equation}
with the initial conditions 
\begin{equation}
\label{eq:hallam_ic}
x_{i_k} (t_0) = \left \{
\begin{aligned}
1, \quad k=0 \\
0, \quad k \neq 0
\end{aligned}
\right.
\nonumber
\end{equation}
for $i = 1,\ldots,3$. There are $n \times (n+1) = 3 \times 4 = 12$ equations in this system.

The system is solved numerically and the graphs for selected elements of the substorage and subthroughflow matrices are depicted in Fig.~\ref{fig:hallam_g1}. Clearly, the substorages and subthroughflows reflect the impact of the unit impulse at about $t=15$. Note that, the system completely recovers after the disturbance in about $10$ time units. This time interval can be taken as a quantitative measure for the {\em restoration time} and {\em system resilience}. As seen from the results, the distribution of environmental nutrient inputs and the organization of the associated nutrient storages generated by the inputs can be analyzed individually and separately within the system. In other words, the evolution of the environmental inputs and associated storages can be tracked individually and separately throughout the system. Therefore, the proposed measures can be used as quantitative ecological indicators for various ecosystem characteristics and behaviors.
\begin{figure}[t]
    \centering
    \begin{subfigure}[b]{0.45\textwidth}
        \includegraphics[width=\textwidth]{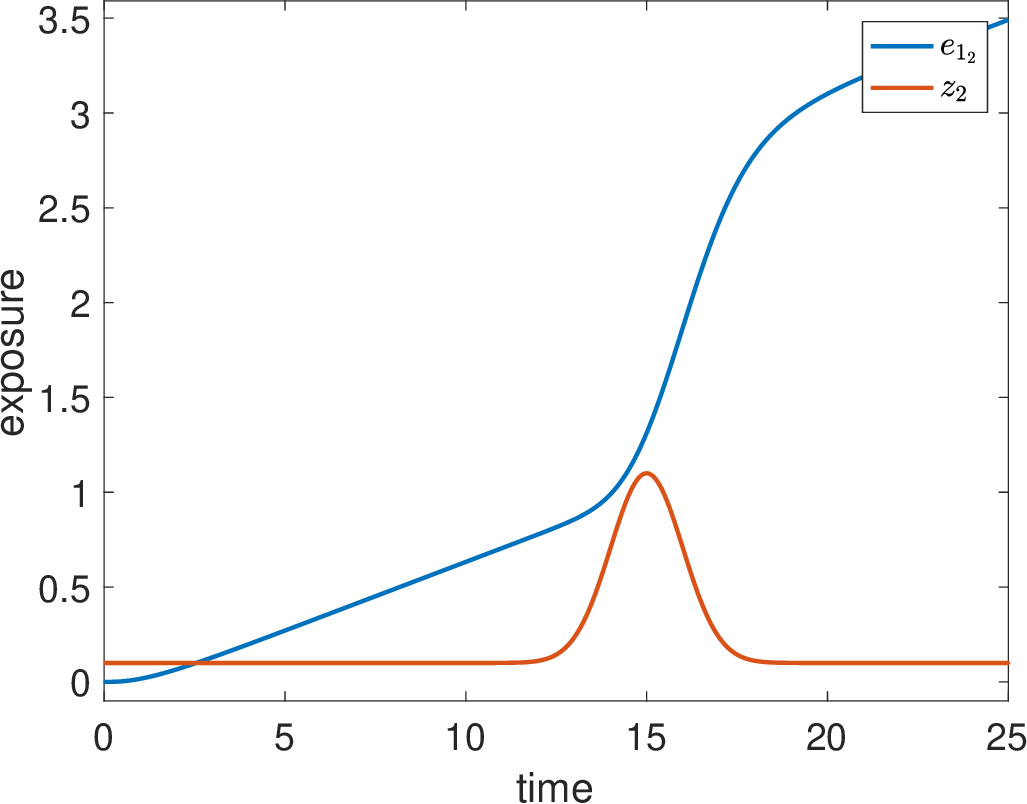}
        \caption{$e_{1_2}(0,t)$ and $z_{2}(t)$}
        \label{fig:hallam_E}
    \end{subfigure}
    \begin{subfigure}[b]{0.45\textwidth}
        \includegraphics[width=\textwidth]{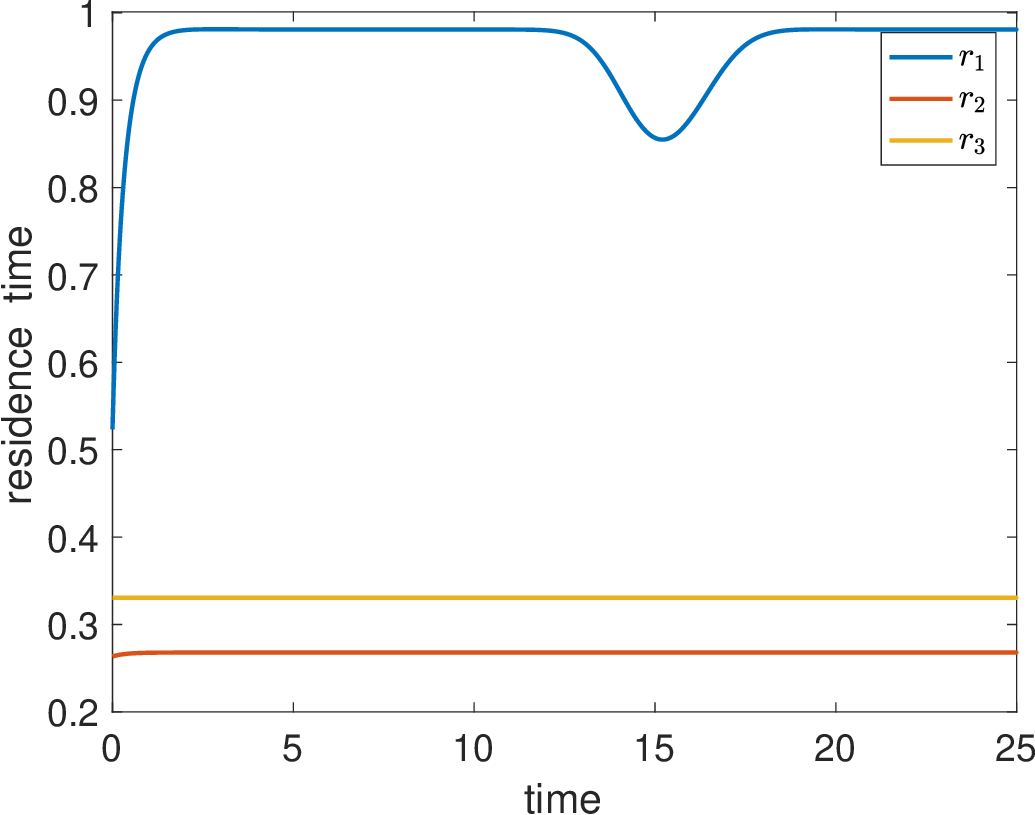}
        \caption{$r_i(t,x)$}
        \label{fig:hallam_R}
    \end{subfigure}
\caption{The graphical representation of (a) the exposure function $e_{1_2}(0,t)$ and the corresponding time-dependent Gaussian impulse $z_{2}(t) = {\mathrm{e}}^{\frac{-(t-15)^2}{2}}+0.1$ and (b) the residence time functions, $r_i(t,x)$ (Case study~\ref{ex:hallam}).}
\label{fig:hallam_rt}
\end{figure}

The \texttt{diact} exposures and residence times are introduced in Section~\ref{sec:EEtime}. The exposure of the resource compartment during the time interval $[5,10]$ to the nutrient input entering the system at the producers compartment, $e_{1_2}(5,10)$, can be obtained as follows:
\begin{equation}
\label{eq:hippe_exp}
\begin{aligned} 
e_{1_2}(5,10) = \int_{5}^{10} x_{1_2}(s) \, ds = 0.36 .
\end{aligned}
\nonumber
\end{equation}
Similarly, $e_{1_2}(20,25) = 0.39$ and $e_{1_2}(12.5,17.5) = 1.81$. As these results indicate, due to the symmetry of the environmental input, $z_2(t)$, the exposure of compartment $1$ to this input during the time interval $[20,25]$ is closer to the exposure to the same input during $[5,10]$. The exposure of the same compartment to the input, however, is greater during the interval $[12.5,17.5]$ about the maximum environmental input at $t=15$. The graph of exposure function $e_{1_2}(0,t)$ is presented together with the corresponding input, $z_2(t)$, for the time interval $[0,25]$ in Fig.~\ref{fig:hallam_E}. There is clearly a sudden increase in the graph of $e_{1_2}(0,t)$ at about $t=15$ due to the disturbance, as expected.

The residence time functions for this model are depicted in Fig.~\ref{fig:hallam_R}. The residence times of both the consumer and producer compartments are almost constant. Interestingly, the Gaussian impulse into the producer compartment, $z_2(t)$, has no significant impact on the activity level of the consumer and even that of the producer compartment itself. However, the maximum impulse at about $t=15$ decreases the residence time of the resource compartment, $r_1(t)$, locally in time. Numerically, 
\[ \mathcal{R}(10) = \mathcal{R}(25) = \diag{([0.98,0.27,0.33])} 
\quad \mbox{but} \quad
\mathcal{R}(15) = \diag{([0.85,0.27,0.33])} . \]
In other words, the residence time of the resource compartment adversely impacted by the environmental input into the producer compartment. Consequently, increasing environmental nutrient input into the producers compartment decreases the residence time of the nutrient storage in and, therefore, increases the activity level of the resource compartment (and all of its subcompartments) only.
\begin{figure}[t]
    \centering
    \begin{subfigure}[b]{0.45\textwidth}
        \includegraphics[width=\textwidth]{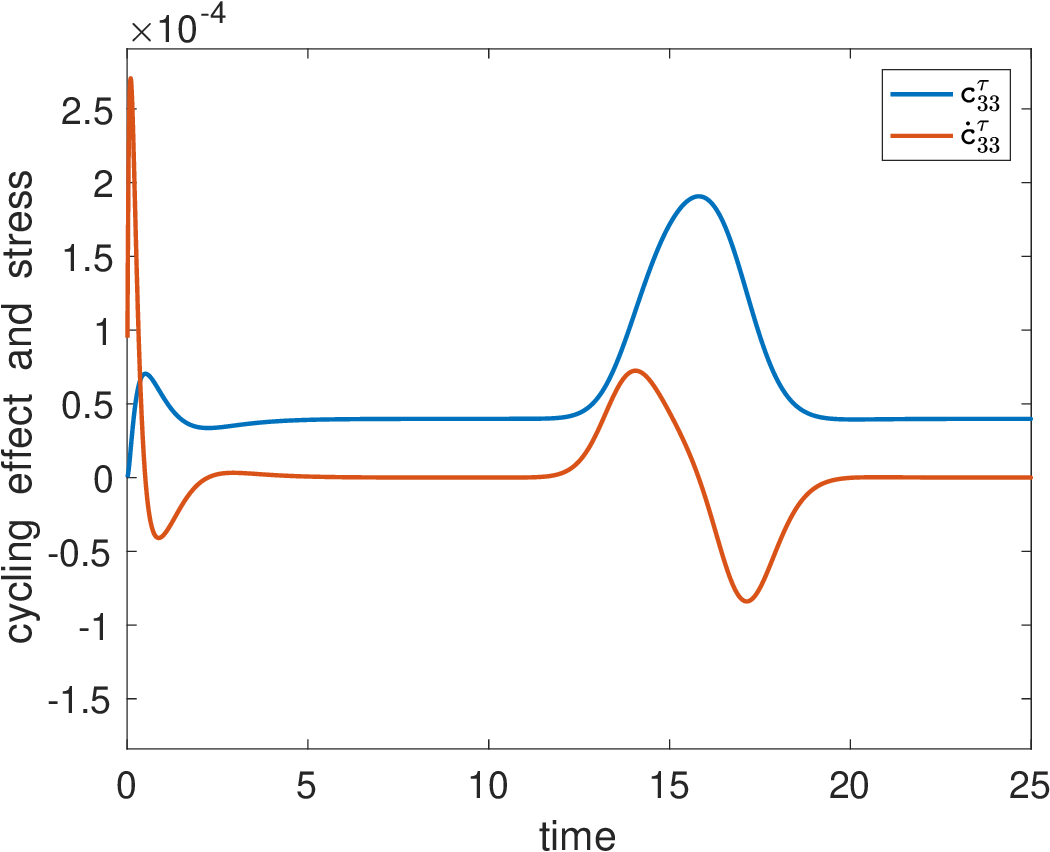}
        \caption{$\texttt{c}_{33}^\tau(t)$ and $\dot{\texttt{c}}_{33}^\tau(t)$}
        \label{fig:hallam_dC}
    \end{subfigure}
    \begin{subfigure}[b]{0.45\textwidth}
        \includegraphics[width=\textwidth]{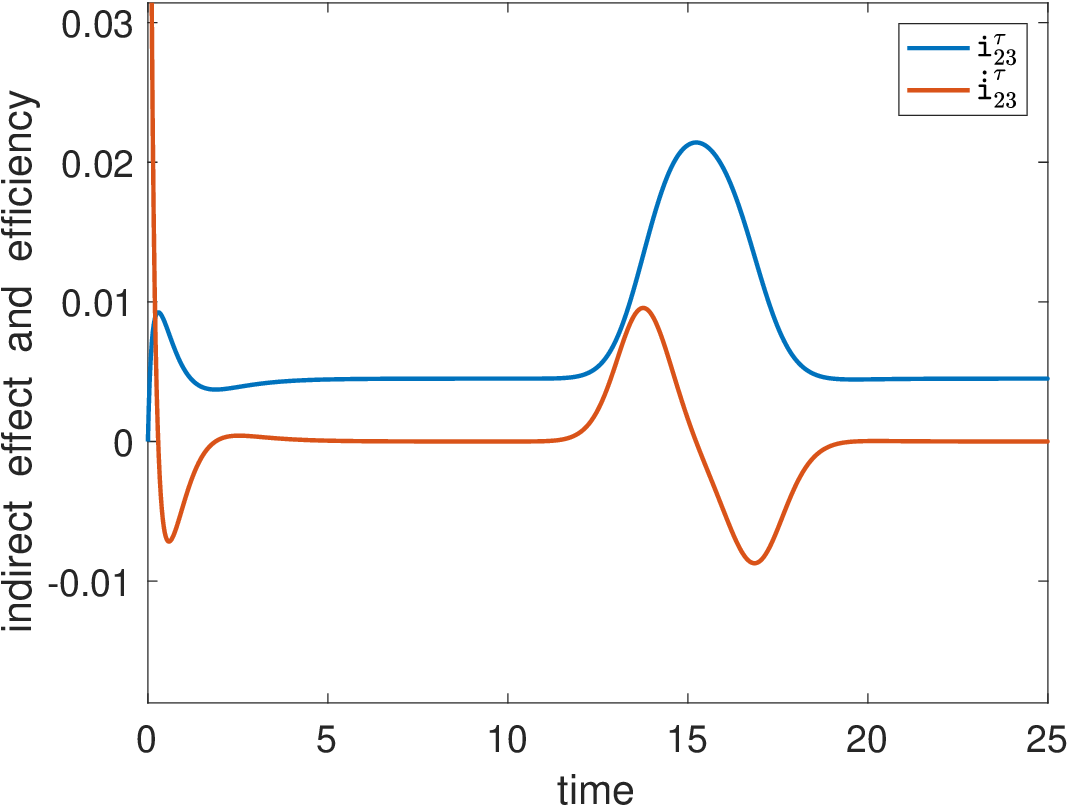}
        \caption{$\texttt{i}_{2 3}^\tau(t)$ and $\dot{\texttt{i}}_{2 3}^\tau(t)$}
        \label{fig:hallam_Ef}
    \end{subfigure}
\caption{The numerical results for (a) the composite cycling effect index of compartment $3$ and the corresponding system stress, $\texttt{c}_{33}^\tau(t)$ and $\dot{\texttt{c}}_{33}^\tau(t)$, and (b) the indirect effect index of compartment $3$ on $2$ and the corresponding efficiency, $\texttt{i}_{2 3}^\tau(t)$ and $\dot{\texttt{i}}_{2 3}^\tau(t)$ (Case study~\ref{ex:hallam}).}
\label{fig:hallam_ind_effs}
\end{figure}

The \texttt{diact} effect measures and indices are introduced in Section~\ref{sec:effects} and the \texttt{diact} flows are listed componentwise in Eq.~\ref{eq:comp_diact_subs}. The unit impulse also manifested itself as rapid fluctuations around the maximum stimulus time, $t=15$, in the graphs of the composite indirect and cycling effect indices, $\texttt{i}^\tau_{2 3}(t) = {\tau^\texttt{i}_{2 3}}(t)/ \check{\sigma}^\tau(t)$ and $\texttt{c}^\tau_{33}(t) = {\tau_{33}^\texttt{c} }(t)/ \check{\sigma}^\tau(t)$, as well as those of the corresponding indirect effect efficiency and stress, as presented in Fig.~\ref{fig:hallam_ind_effs}. Although they have different values, both indices have the same behavior due to their complementary nature \cite{Coskun2017DCSAM}. The unusual rapid fluctuations in the effect efficiency and stress indicate an excess amount of nutrient input into the system. Therefore, they can be used to quantify the system resilience and resistance to disturbances as discussed in Section~\ref{sec:effects}. The maximum period of these fluctuations can be used as a measure for the system resilience, similar to the subthroughflows and substorages as discussed above. The maximum amplitude of the stress can be used as a measure for the system resistance. Using this indirect effect efficiency and stress, $\dot{\texttt{i}}^\tau_{2 3}(t)$ and $\dot{\texttt{c}}^\tau_{33}(t)$, the system resilience is about 10 time units, and the system resistance is approximately $10^{-4}$. Consequently, the system efficiency and stress as ecological indicators can monitor ecosystems for environmental impacts.
\begin{figure}[t]
    \centering
    \begin{subfigure}[b]{0.31\textwidth}
        \includegraphics[width=\textwidth]{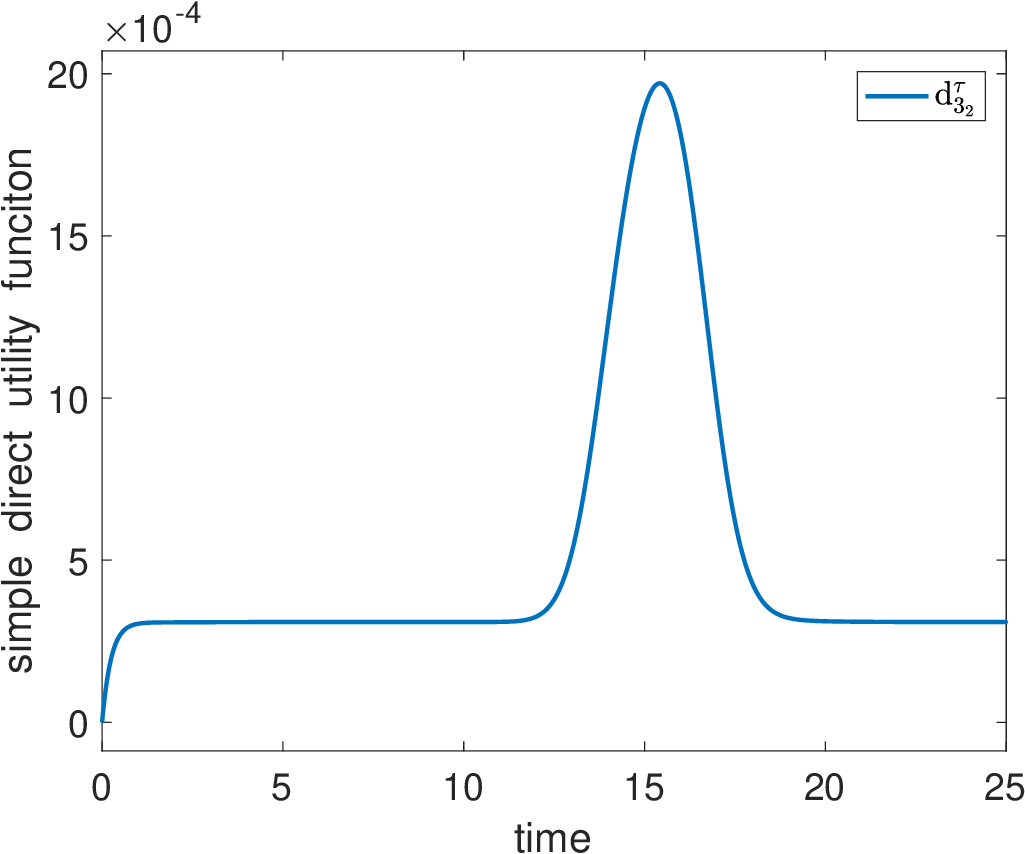}
        \caption{$\mathbbm{d}^{\tau}_{3_2}(t)$}
        \label{fig:hallam_u}
    \end{subfigure}
    \begin{subfigure}[b]{0.32\textwidth}
        \includegraphics[width=\textwidth]{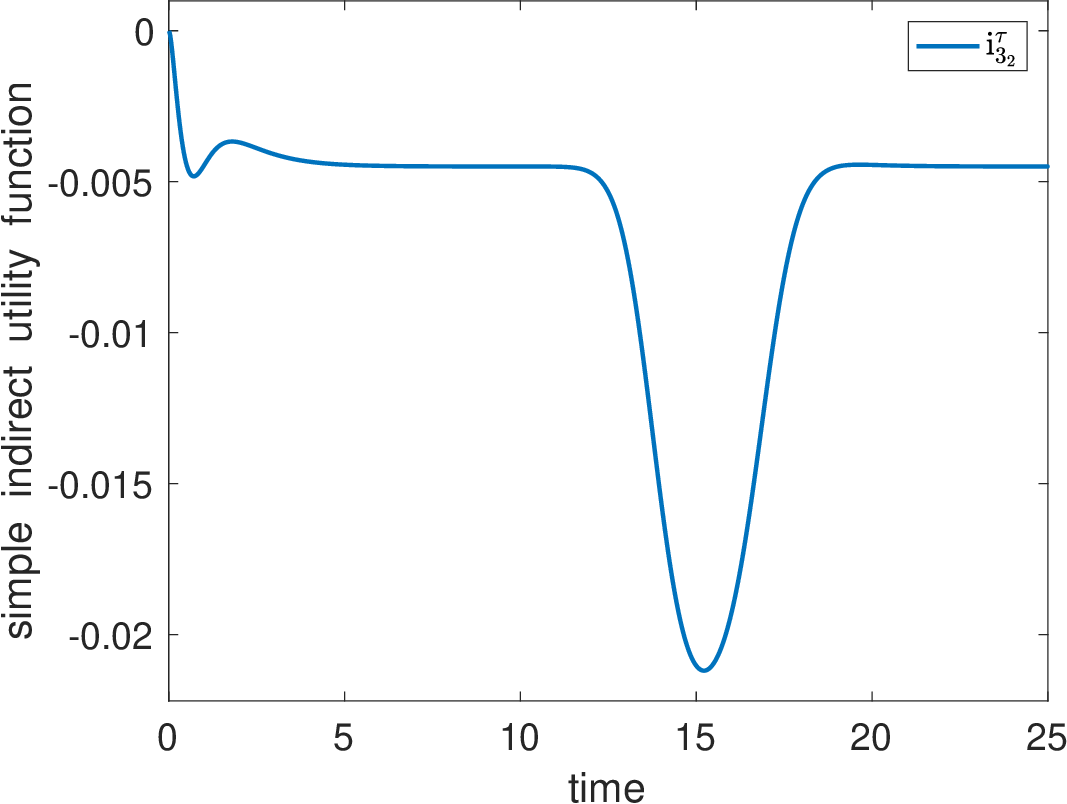}
        \caption{$\mathbbm{i}^{\tau}_{3_2}(t)$}
        \label{fig:hallam_uu}
    \end{subfigure}
    \begin{subfigure}[b]{0.32\textwidth}
        \includegraphics[width=\textwidth]{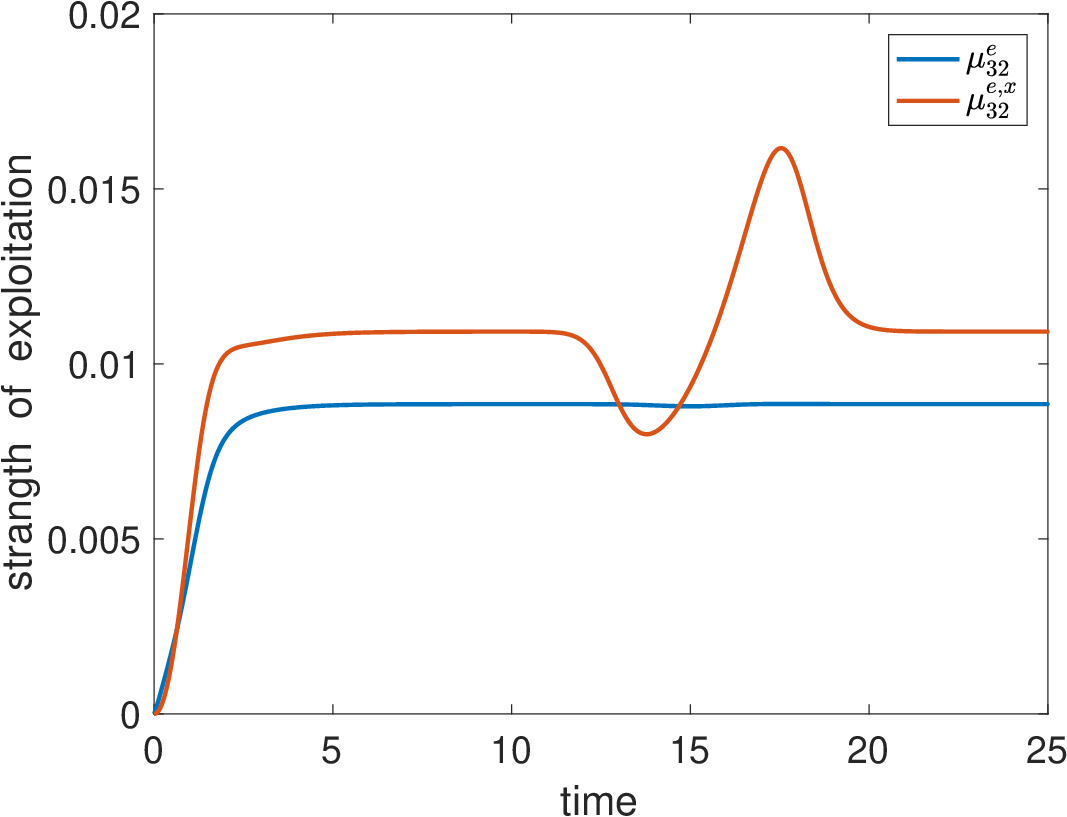}
        \caption{$\mu_{32}^e(t)$ and $\mu_{32}^{e,x}(t)$}
        \label{fig:hallam_intint}
    \end{subfigure}
\caption{The numerical results for (a) the flow-based simple direct and (b) indirect utility indices of compartment $2$ to $3$, $\mathbbm{d}^{\tau}_{3_2}(t)$ and $\mathbbm{i}^{\tau}_{3_2}(t)$, induced only by the corresponding inputs $z_2(t)$ and $z_3(t)$, and (c) the strength of the exploitation relationship between these compartments. In the figure legends, $\mathrm{d}$ and $\mathrm{i}$ notations are used for $\mathbbm{d}$ and $\mathbbm{i}$ due to the limited font library of Matlab software (Case study~\ref{ex:hallam}). }
\label{fig:hallam_g2}
\end{figure}

The \texttt{diact} utility measures and indices are introduced in Section~\ref{sec:utility}. The flow-based simple transfer utility index induced only by inputs $z_k(t)$ and $z_i(t)$ can be expressed as $\mathbbm{t}^{\tau}_{i_k}(t) = \texttt{t}^\tau_{i_k} -  \texttt{t}^\tau_{k_i}$ where the corresponding simple transfer effect index of compartment $k$ on $i$ induced only by $z_k(t)$ is formulated as $\texttt{t}^\tau_{i_k}(t) = \tau^{\texttt{t}}_{i_k k_k}(t) / {\check{\sigma}^\tau(t)} = \tau^{\texttt{t}}_{i_k}(t) / {\check{\sigma}^\tau(t)}$. The flow-based simple direct and indirect utilities transmitted from the producers to the consumers induced only by the corresponding nutrient inputs, $z_2(t)$ and $z_3(t)$, can then be formulated as follows:
\begin{equation}
\label{eq:rest_diact}
\begin{aligned} 
\mathbbm{d}^{\tau}_{3_2}(t) & = \texttt{d}^\tau_{3_2}(t) -  \texttt{d}^\tau_{2_3}(t) 
= \frac{ \tau^\texttt{d}_{3_2}(t) - \tau^\texttt{d}_{2_3}(t) } {\check{\sigma}^\tau(t) }
= \frac{ \tau^\texttt{d}_{3_2 2_2}(t) - \tau^\texttt{d}_{2_3 3_3}(t) } {\check{\sigma}^\tau(t) } \\
\mathbbm{i}^{\tau}_{3_2}(t) & = \texttt{i}^\tau_{3_2}(t) -  \texttt{i}^\tau_{2_3}(t) 
= \frac{ \tau^\texttt{i}_{3_2}(t) - \tau^\texttt{i}_{2_3}(t) } {\check{\sigma}^\tau(t) } 
= \frac{ \tau^\texttt{i}_{3_2 2_2}(t) - \tau^\texttt{i}_{2_3 3_3}(t) } {\check{\sigma}^\tau(t) } 
\nonumber 
\end{aligned} 
\end{equation}
The direct and indirect subflows in the expressions above are computed as formulated in Eq.~\ref{eq:comp_diact_subs}. The graphical representation of these functions are given in Fig.~\ref{fig:hallam_g2}. Both graphs have fluctuations due to the Gaussian impulse at about $t=15$. Interestingly, while the simple direct utility function is always positive, $\mathbbm{d}^{\tau}_{3_2}(t)>0$, the simple indirect utility function is negative, ${\mathbbm{i}}^{\tau}_{3_2}(t)<0$, during $[0,25]$. That is, considering the effects induced only by nutrient inputs $z_2(t)$ and $z_3(t)$, although the consumer compartment has relative nutrient gain (benefit) from the producer compartment through direct interactions, it has relative nutrient loss to (harm from) the producer compartment indirectly through the resource compartment.

The quantitative definitions of the main types of interspecific interactions are introduced in Section~\ref{sec:qdii}. Based on these definitions, there is no neutralism, mutualism, commensalism, or competition in this ecosystem. Disregarding the resource compartment and since $\texttt{d}^\tau_{32}(t) > 0$ and $\texttt{d}^\tau_{23}(t) = 0$, the only interspecific interaction exists in this system is exploitation between the producer and consumer compartments. The flow- and storage-based strength of the ``predation'' of the consumer compartment on the producer, $\mu_{32}^e(t) = {\tau^\texttt{d}_{32}(t)} / {\tau_{2}(t)}$ and $\mu_{32}^{e,x}(t) = {x^\texttt{d}_{32}(t)} / {x_{2}(t)}$, are presented in Fig.~\ref{fig:hallam_intint}.

\subsection{Case study}
\label{appsec:ex_hippe}

A linear dynamic ecosystem model introduced by \cite{Hippe1983} was recently analyzed through the system decomposition~\cite{Coskun2017DCSAM}. In particular, analytic solutions for the substorages, subthroughflows, as well as the transient and \texttt{diact} flows and storages are presented for this model. In this case study, we present some of the measures and indices introduced in this article for the model.

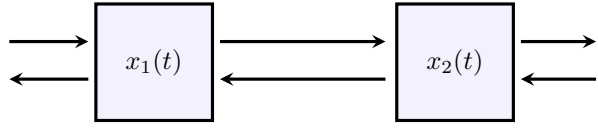
\begin{figure}[t]
\begin{center}
\begin{tikzpicture}
\centering
   \draw[very thick,  fill=blue!5, draw=black] (-.05,-.05) rectangle node(R1) {$x_1(t)$} (1.5,1.5) ;
   \draw[very thick,  fill=blue!5, draw=black] (3.95,-.05) rectangle node(R2) {$x_2(t)$} (5.5,1.5) ;   
       \draw[very thick,stealth-,draw=black]  (1.6,.5) -- (3.8,.5) ;
       \draw[very thick,-stealth,draw=black]  (1.6,1) -- (3.8,1) ;  
       \draw[very thick,stealth-,draw=black]  (5.6,.5) -- (6.6,.5) ;
       \draw[very thick,-stealth,draw=black]  (5.6,1) -- (6.6,1) ;  
       \draw[very thick,stealth-,draw=black]  (-1.2,.5) -- (-0.15,.5) ;
       \draw[very thick,-stealth,draw=black]  (-1.2,1) -- (-0.15,1) ;  
\end{tikzpicture}
\end{center}
\caption{Schematic representation of the model network (Case study~\ref{appsec:ex_hippe}).}
\label{fig:hippe_diag}
\end{figure}

The model has two compartments, $x_1(t)$ and $x_2(t)$ (see Fig.~\ref{fig:hippe_diag}). The flows regime for the system is described as
\begin{equation}
\label{eq:hippe_flows}
\begin{aligned} 
F(t,x) = 
\begin{bmatrix}
    0  & \frac{2}{3} x_2(t)  \\
    \frac{4}{3} x_1(t)  & 0 \\
\end{bmatrix},
\quad
z(t,x) = 
\begin{bmatrix}
    z_1(t) \\
    z_2(t) 
\end{bmatrix},
\quad 
y(t,x) = 
\begin{bmatrix}
    \frac{1}{3} x_1(t) \\
    \frac{5}{3} x_2(t)
\end{bmatrix} .
\nonumber
\end{aligned}
\end{equation}
The governing equations take the following form:
\begin{equation}
\label{eq:hippe_model}
\begin{aligned} 
\dot x_1(t) & = z_1(t) + \frac{2}{3} x_2(t) - \left(\frac{4}{3} + \frac{1}{3} \right) \, x_1(t) \\
\dot x_2(t) & = z_2(t) + \frac{4}{3} x_1(t) - \left(\frac{2}{3} + \frac{5}{3} \right) \, x_2(t)
\end{aligned} 
\nonumber
\end{equation}
with the initial conditions $[x_{1,0}, x_{2,0}]^T = [3,3]^T$.

The subcompartmentalization step yields the substorages as follows:
\[ x_{1_k}(t) \quad \mbox{and} \quad x_{2_k}(t)
\quad \mbox{with} \quad 
x_i(t) = \sum_{k=0}^2 x_{i_k} (t) . \]
The flow partitioning then yields the subflows for the subsystems:
\begin{equation}
\label{eq:hippe_flows_SC}
\begin{aligned} 
F_k(t,{\rm x}) = 
\begin{bmatrix}
    0  & \frac{2}{3} \, d_{2_k} \, x_2 \\
    \frac{4}{3} d_{1_k} \, x_1  & 0 \\
\end{bmatrix},
\quad
\check{z}_k(t,{\rm x}) = 
\begin{bmatrix}
    \delta_{1k} \, z_1 \\
    \delta_{2k} \, z_2
\end{bmatrix},
\quad 
\hat{y}_k(t,{\rm x}) = 
\begin{bmatrix}
    \frac{1}{3} d_{1_k} \, x_1 \\
    \frac{5}{3} \, d_{2_k} \, x_2
\end{bmatrix} ,
\end{aligned}
\nonumber
\end{equation}
where the decomposition factors, $d_{i_k}({\rm x})$, are defined by Eq.~\ref{eq:cons2_1new}. Consequently, the dynamic system partitioning methodology yields the following decomposed system:
\begin{equation}
\label{eq:hippe_sc}
\begin{aligned} 
\dot x_{1_k}(t) & = z_{1_k}(t) + \frac{2}{3} x_{2_k}(t) - \left(\frac{4}{3} + \frac{1}{3} \right) \, x_{1_k}(t) \\
\dot x_{2_k}(t) & = z_{2_k}(t) + \frac{4}{3} x_{1_k}(t) - \left(\frac{2}{3} + \frac{5}{3} \right) \, x_{2_k}(t)
\end{aligned}
\nonumber
\end{equation}
with the initial conditions 
\begin{equation}
\label{eq:hippe_ic}
x_{i_k} (t_0) = \left \{
\begin{aligned}
3, \quad k=0 \\
0, \quad k \neq 0
\end{aligned}
\right.
\nonumber
\end{equation}
for $i = 1,2$. There are $n \times (n+1) = 2 \times 3 = 6$ equations\textemdash one for each subcompartment.
\begin{figure}[t]
    \centering
    \begin{subfigure}[b]{0.44\textwidth}
        \includegraphics[width=\textwidth]{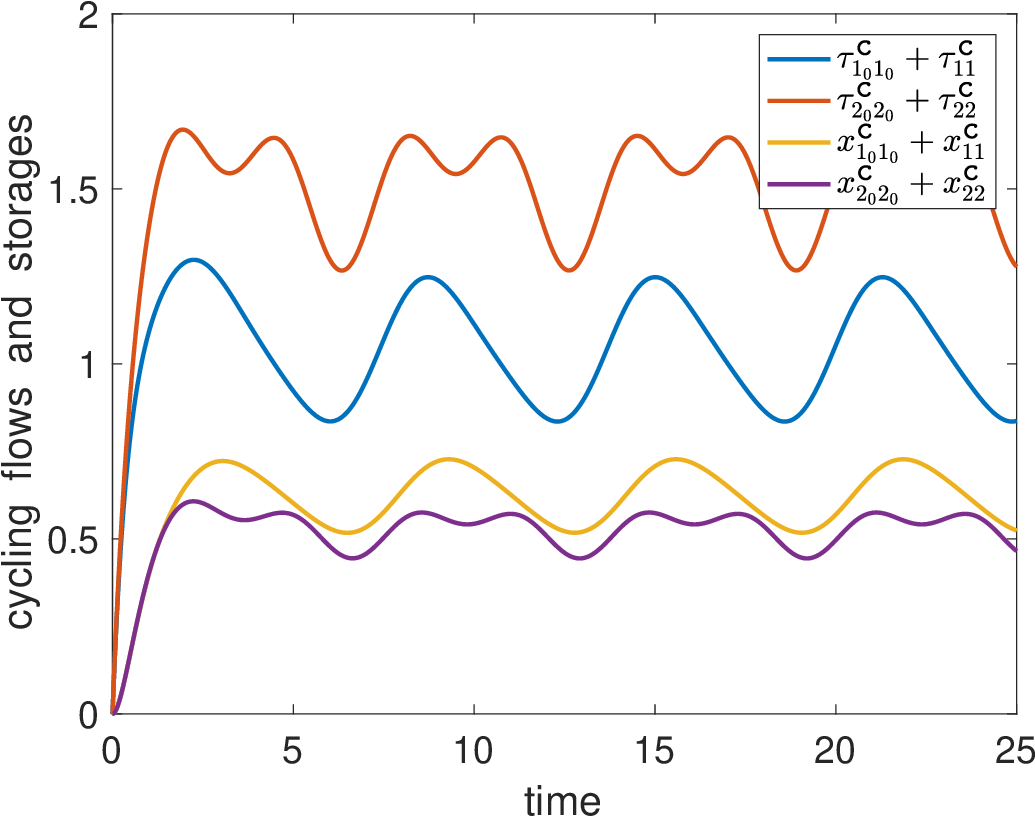}
        \caption{$\tau^\texttt{c}_{i_0 i_0}(t) + \tau^\texttt{c}_{ii}(t)$ and $x^\texttt{c}_{i_0 i_0}(t) + x^\texttt{c}_{ii}(t)$}
        \label{fig:hippe_cyc_sub}
    \end{subfigure}
    \begin{subfigure}[b]{0.45\textwidth}
        \includegraphics[width=\textwidth]{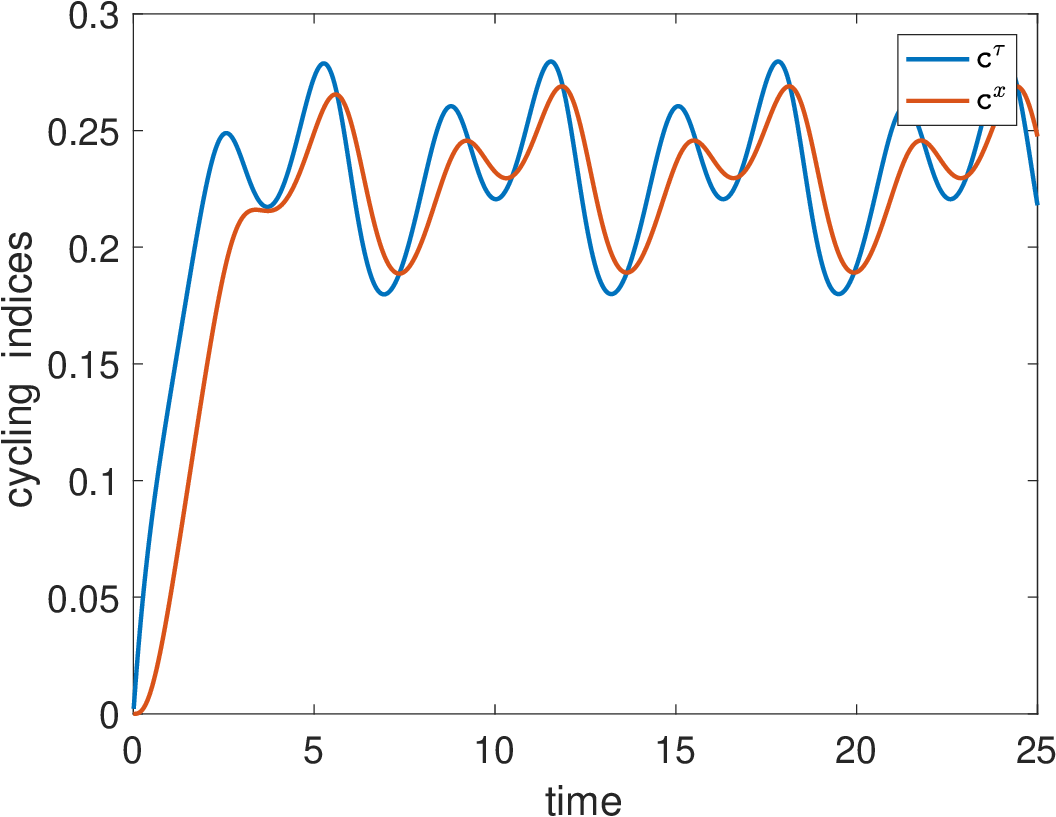}
        \caption{$\tilde{\texttt{c}}^\tau(t)$ and $\tilde{\texttt{c}}^x(t)$}
        \label{fig:hippe_cyci}
    \end{subfigure}
\caption{The graphical representation of (a) the composite cycling flows and storages, $\tau^\texttt{c}_{i_0 i_0}(t) + \tau^\texttt{c}_{ii}(t)$ and $x^\texttt{c}_{i_0 i_0}(t) + x^\texttt{c}_{ii}(t)$, and (b) flow- and storage-based simple cycling indices, $\tilde{\texttt{c}}^\tau(t)$ and $\tilde{\texttt{c}}^x(t)$ (Case study~\ref{appsec:ex_hippe}).}
\label{fig:hippe_cyc}
\end{figure}

The system is solved analytically with a time-dependent input of $z(t) = [3+\sin(t),3+\sin(2 t)]^T$. Some elements of the substorage and subthroughflow matrix functions are
\begin{equation}
\label{eq:hippe_pedX}
\begin{aligned} 
x_{1_1}(t) & = \frac{7}{3} - \frac{11\,\cos\left(t\right)}{30} + \frac{13\,\sin\left(t\right)}{30} -\frac{5\,{\mathrm{e}}^{-t}}{3} - \frac{3\,{\mathrm{e}}^{-3\,t}}{10} \\
\check{\tau}_{1_2}(t) & = \frac{742}{585} - \frac{184\, \cos^2\left(t\right)}{585} +\frac{86\,\sin\left(2\,t\right)}{585} -\frac{26\,{\mathrm{e}}^{-t}}{45}-\frac{44\,{\mathrm{e}}^{-3\,t}}{117} 
\end{aligned} 
\nonumber
\end{equation}  
as given in \cite{Coskun2017SCSA}. 

The composite cycling subflows can be obtained by using the formulations in Eq.~\ref{eq:comp_diact_subs}. Analytically, $\tau^{\texttt{c}}_{1_0 1_0}(t)$, for example, can be expressed as 
\begin{equation}
\label{eq:cyc_analytic}
\begin{aligned}
\tau^{\texttt{c}}_{1_0 1_0}(t) = -\frac{ 36 \, {\mathrm{e}}^{-t} -100\,{\mathrm{e}}^{t} + 80\,{\mathrm{e}}^{2\,t} -16\,{\mathrm{e}}^{2\,t}\,\cos\left(t\right)+8\,{\mathrm{e}}^{2\,t}\,\sin\left(t\right) }
{9 + 50\,{\mathrm{e}}^{2\,t}-70\,{\mathrm{e}}^{3\,t}+11\,{\mathrm{e}}^{3\,t}\,\cos\left(t\right)-13\,{\mathrm{e}}^{3\,t}\,\sin\left(t\right) } . 
\end{aligned}
\end{equation}
The composite cycling substorages can then be computed by coupling Eq.~\ref{eq:out_in_diact2} for the cycling subflows and substorages with the decomposed system, Eq.~\ref{eq:model_M}, and solving them simultaneously. Since the model is linear, Eq.~\ref{eq:out_in_diact2} can be solved analytically as well \cite{Coskun2017DCSAM}. The graphs of the composite cycling flows and storages induced both by the environmental inputs and initial stocks,
\begin{equation}
\label{eq:cyc2_sln}
\begin{aligned}
{\tau}^\texttt{c}_{i_0 i_0}(t) + {\tau}^\texttt{c}_{ii}(t) = \sum_{k=0}^2 {\tau}^\texttt{c}_{i_k i_k}(t)  \quad & \mbox{and} \quad x^\texttt{c}_{i_0 i_0}(t) + x^\texttt{c}_{ii}(t) = \sum_{k=0}^2 x^\texttt{c}_{i_k i_k}(t) 
\end{aligned}
\end{equation}
for $i=1,2$, are presented in Fig.~\ref{fig:hippe_cyc_sub}.

The flow- and storage-based simple cycling effect indices induced by environmental inputs can then be expressed for the system as defined in Eq.~\ref{eq:diact_eff}:
\begin{equation}
\label{eq:cyc_index}
\begin{aligned}
\tilde {\texttt{c}}^\tau(t) = \frac{ {\tau}^\texttt{c}_{11}(t) + {\tau}^\texttt{c}_{22}(t) }{ \check{\tau}_{1}(t) + \check{\tau}_{2}(t) } \quad \mbox{and} \quad \tilde{\texttt{c}}^x(t) = \frac{ {x}^\texttt{c}_{11}(t) + {x}^\texttt{c}_{22}(t) }{ x_{1}(t) + x_{2}(t) } .
\end{aligned}
\end{equation}
Their graphs are presented in Fig.~\ref{fig:hippe_cyci}. As seen from the graphs, the flow- and storage-based cycling effect indices have a similar behavior. Due to the periodic behavior of the environmental inputs, the cycling effect indices induced by the environmental inputs are also periodic.

The residence time matrix for this model, defined in Eq.~\ref{eq:irestimeM}, becomes
\begin{equation}
\label{eq:hippe_res} 
\begin{aligned} 
\mathcal{R}(t,{x}) = \diag{( [0.6, 0.43] )} .
\end{aligned} 
\nonumber
\end{equation} 
The residence time of compartment $2$ is constantly smaller than that of compartment 1. That is, $r_2(t,x) = 0.43 < 0.6 = r_1(t,x)$. This result can ecologically be interpreted as compartment $2$ being more active, in terms of storage transfer, than compartment $1$.

\subsection{Case study}
\label{ex:disc}
\begin{figure}[t]
\begin{center}
\begin{tikzpicture}
\centering
   \draw[very thick,  fill=blue!5, draw=black] (-.05,-.05) rectangle node(R1) {$x_1$} (1.5,1.5) ;
   \draw[very thick,  fill=blue!5, draw=black] (3.95,-.05) rectangle node(R2) {$x_2$} (5.5,1.5) ;   
   \draw[very thick,  fill=blue!5, draw=black] (-.05,3.95) rectangle node(R3) {$x_4$} (1.5,5.5) ;      
   \draw[very thick,  fill=blue!5, draw=black] (3.95,3.95) rectangle node(R3) {$x_5$} (5.5,5.5) ;      
   \draw[very thick,  fill=blue!5, draw=black] (7.95,-.05) rectangle node(R3) {$x_7$} (9.5,1.5) ;      
   \draw[very thick,  fill=blue!5, draw=black] (7.95,3.95) rectangle node(R3) {$x_6$} (9.5,5.5) ;      
   \draw[very thick,  fill=blue!5, draw=black] (11.95,2) rectangle node(R3) {$x_3$} (13.5,3.5) ;      
       \draw[very thick,stealth-,draw=black]  (4.9,-1) -- (4.9,-.15) ;     
       \draw[very thick,-stealth,draw=black]  (4.6,-1) -- (4.6,-.15) ;       
       \draw[very thick,stealth-,draw=black]  (.9,-1) -- (.9,-.15) ;    
       \draw[very thick,-stealth,draw=black]  (.6,-1) -- (.6,-.15) ;      
       \draw[very thick,stealth-,draw=black]  (8.9,-1) -- (8.9,-.15) ; 
       \draw[very thick,-stealth,draw=black]  (8.6,-1) -- (8.6,-.15) ;     
       \draw[very thick,stealth-,draw=black]  (4.9,6.5) -- (4.9,5.6) ;     
       \draw[very thick,-stealth,draw=black]  (4.6,6.5) -- (4.6,5.6) ;       
       \draw[very thick,stealth-,draw=black]  (.9,6.5) -- (.9,5.6) ;    
       \draw[very thick,-stealth,draw=black]  (.6,6.5) -- (.6,5.6) ;      
       \draw[very thick,stealth-,draw=black]  (8.9,6.5) -- (8.9,5.6) ; 
       \draw[very thick,-stealth,draw=black]  (8.6,6.5) -- (8.6,5.6) ;     
       \draw[very thick,-stealth,draw=black]  (12.8,1) -- (12.8,1.9) ;     
       \draw[very thick,-stealth,draw=black]  (12.8,3.6) -- (12.8,4.6) ;     
       \draw[very thick,-stealth,draw=black]  (.75,1.6) -- (.75,2.7) -- (11.9,2.7) ;    
       \draw[very thick,stealth-stealth,draw=black]  (5.2,1.6) -- (5.2,2.6) -- (11.9,2.6) ;       
       \draw[very thick,stealth-stealth,draw=black]  (5.2,3.9) -- (5.2,2.8) -- (11.9,2.8) ;  
       \draw[very thick,stealth-stealth,draw=black]  (8.75,3.9) -- (8.75,2.9) -- (11.9,2.9) ;       
       \draw[very thick,-stealth,draw=black]  (8.75,1.6) -- (8.75,2.5) -- (11.9,2.5) ;  
       \draw[very thick,-stealth,draw=black]  (1.6,.8) -- (3.9,.8) ;  
       \draw[very thick,-stealth,draw=black]  (1.6,1.4) -- (4,2.3) -- (5.6,2.3) -- (8.6,1.6) ;  
       \draw[very thick,stealth-,draw=black]  (1.2,1.6) -- (7.8,3.9) ;  
       \draw[very thick,stealth-,draw=black]  (.9,1.6) -- (4.4,3.8) ;  
       \draw[very thick,stealth-stealth,draw=black]  (.6,1.6) -- (.6,3.8) ;  
       \draw[very thick,stealth-stealth,draw=black]  (5.6,.8) -- (7.9,.8) ;  
       \draw[very thick,stealth-stealth,draw=black]  (4,1.6) -- (1,3.8) ;  
       \draw[very thick,stealth-,draw=black]  (4.8,1.6) -- (4.8,3.8) ;  
       \draw[very thick,stealth-stealth,draw=black]  (5.5,1.6) -- (8.5,3.8) ;  
       \draw[very thick,stealth-,draw=black]  (5.6,4.7) -- (7.9,4.7) ;  
       \end{tikzpicture}
\end{center}
\caption{Schematic representation of the model network (Case Study~\ref{ex:disc}).}
\label{fig:disc_diag}
\end{figure}
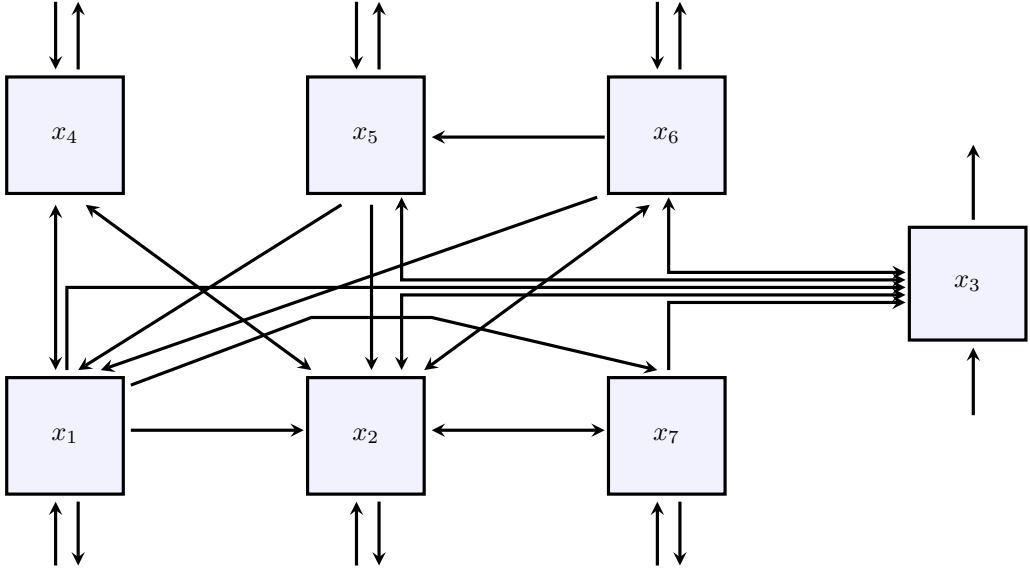

The Neuse River Estuary is a drowned river valley located at the transition from the Neuse River to Pamlico Sound in North Carolina. In 1997, the State of North Carolina legislated a reduction in nitrogen loading to the estuary. As part of the monitoring program to study the estuary's response to new environmental management, nitrogen loading data is constructed for 16 seasons starting from Spring 1985 to Winter 1989 \cite{Christian2000}.

The Neuse River Estuary ecosystem is modeled with seven compartments: phytoplankton particulate nitrogen, $1-$PN-phyto; heterotroph particulate nitrogen, $2-$PN-hetero; sediment particulate nitrogen, $3-$N-sed; dissolved organic nitrogen, $4-$DON; nitrate and nitrites, $5-$NOx; ammonium, $6-$NH$4$; and abiotic particulate nitrogen, $7-$PN-abiotic. The conserved quantity of interest in this case is nitrogen. The compartments are indexed in the given order; for example, $x_1(t)$ represents the nitrogen storage in PN-phyto at time $t$ (see Fig.~\ref{fig:disc_diag}). The units for nitrogen storage and flow are (mmol m$^{-2}$) and (mmol m$^{-2}$ season$^{-1}$), respectively. Each season is considered to be a discrete time step; for example, $t=1$ corresponds to Spring 1985 and $t=16$ to Winter 1989. At each time step, the system is at steady state. 
\begin{figure}[t]
    \centering
    \begin{subfigure}[b]{0.44\textwidth}
        \includegraphics[width=\textwidth]{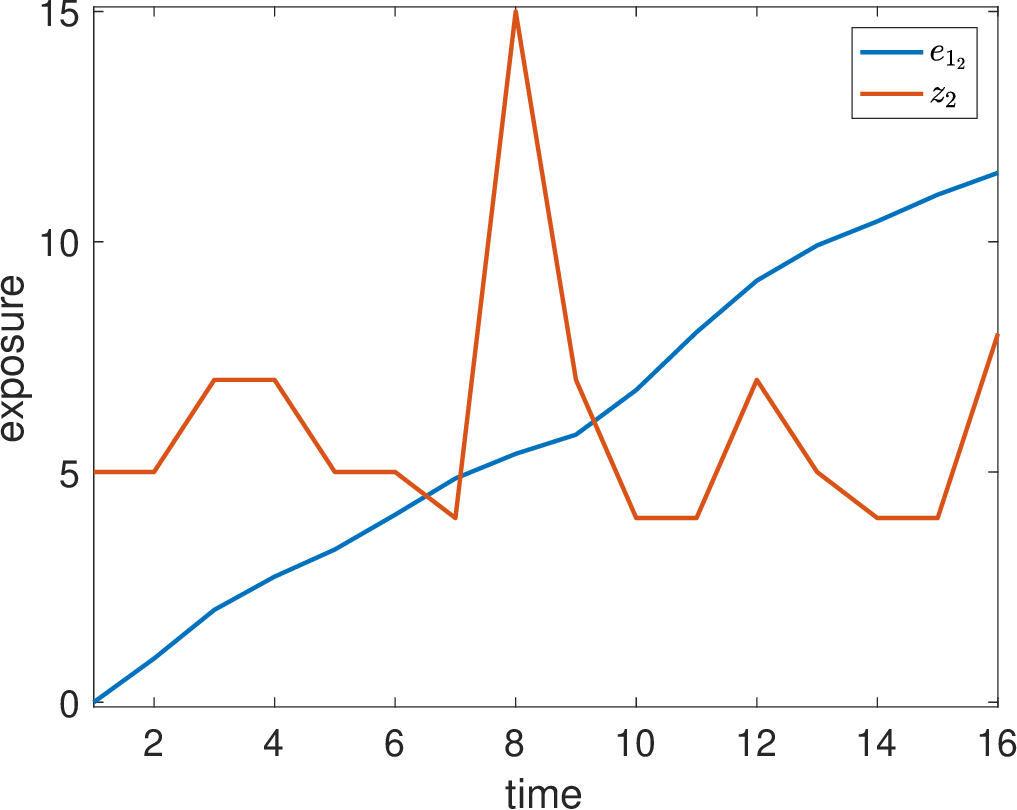}
        \caption{$e_{1_2}(t)$ and $z_{2}(t)$}
        \label{fig:real_expose}
    \end{subfigure}
    \begin{subfigure}[b]{0.45\textwidth}
        \includegraphics[width=\textwidth]{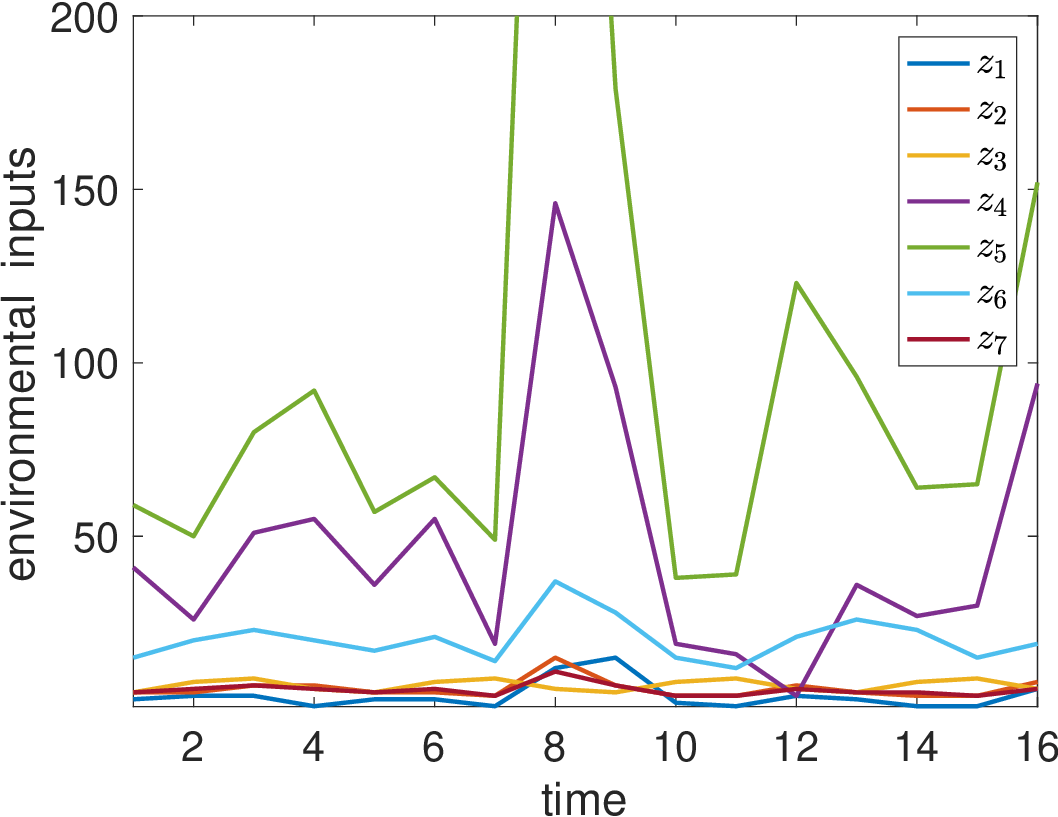}
        \caption{$z_i(t)$}
        \label{fig:real_z}
    \end{subfigure} \\
    \begin{subfigure}[b]{0.44\textwidth}
        \includegraphics[width=\textwidth]{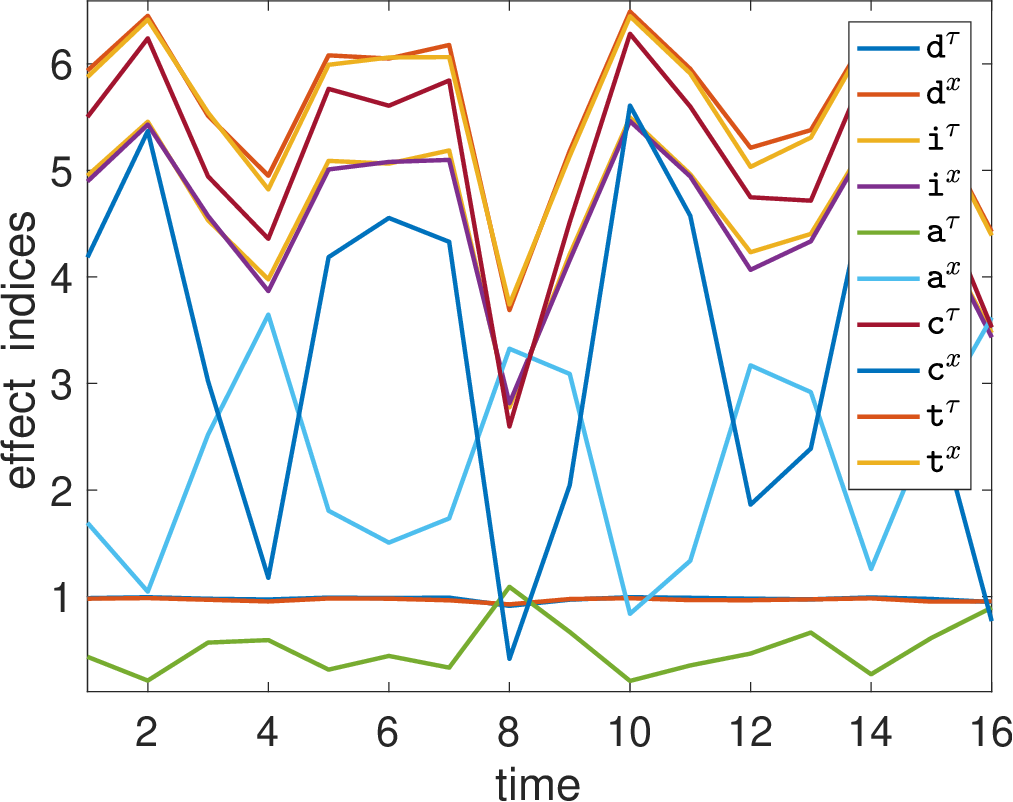}
        \caption{system \texttt{diact} effect indices}
        \label{fig:real_eff}
    \end{subfigure}
    \begin{subfigure}[b]{0.45\textwidth}
        \includegraphics[width=\textwidth]{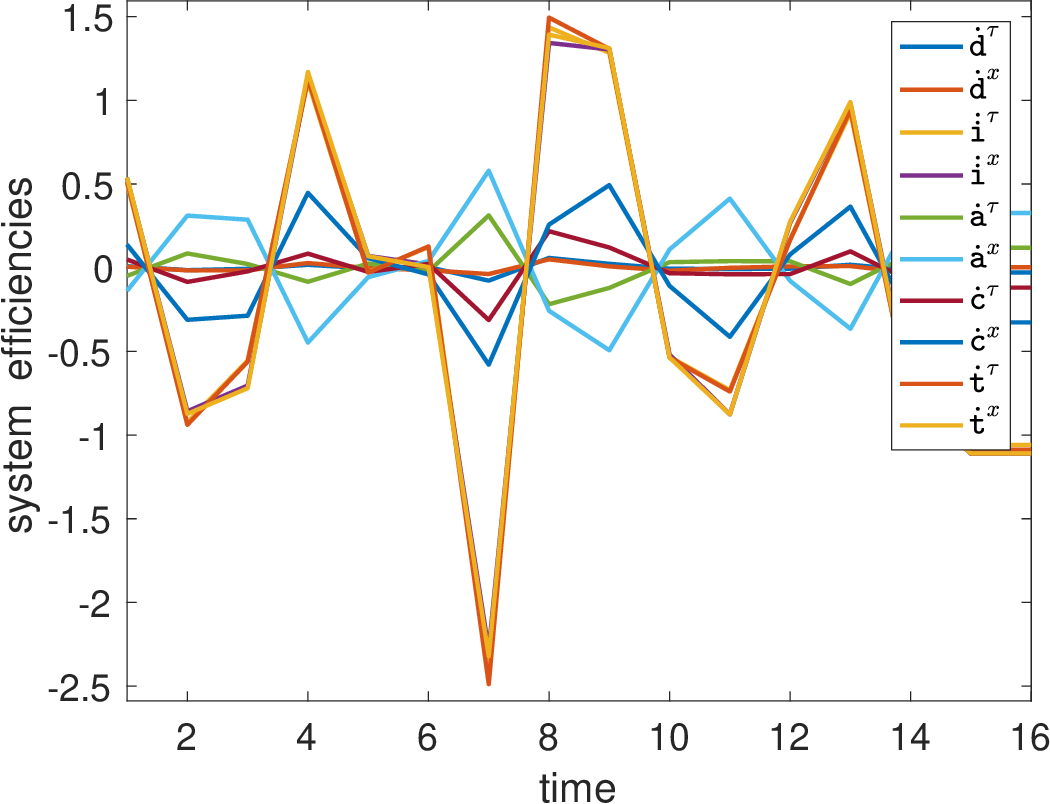}
        \caption{system \texttt{diact} efficiencies}
        \label{fig:real_seff}
    \end{subfigure}    
\caption{The numerical results for (a) the exposure and corresponding environmental input, $e_{1_2}(t)$ and $z_{2}(t)$, (b) the environmental inputs, $z_i(t)$, (c) the system \texttt{diact} effect indices, (d) and the corresponding \texttt{diact} efficiencies (Case study~\ref{ex:disc}).}
\label{fig:disc_ind2}
\end{figure}

The Neuse River Estuary ecosystem model was recently analyzed through the system decomposition theory~\cite{Coskun2017SCSA,Coskun2017SESM}. The subthroughflow and substorage matrix measures, the transient and \texttt{diact} flows and storages, the measures and indices for the \texttt{diact} effects, utilities, and residence times are presented for this ecosystem model in these papers. It has been demonstrated that the proposed method can effectively detect and quantify system properties and phenomena, such as the seasonality, dominance of indirect effects, and high phytoplankton production that the Neuse River and its estuary were experiencing, from the experimental data. This discrete model has been extensively studied in the literature, but some of the results obtained through the system decomposition have not been observed in these studies or could not be demonstrated, although anticipated~\cite{Coskun2017SCSA,Coskun2017SESM}.

The dynamic \texttt{diact} system efficiencies and exposures are introduced in the present paper as the time derivative of the effect indices and the integral of the substorage functions, respectively. The numerical results for the \emph{discrete versions} of these dynamic measures and indices are presented for the Neuse River Estuary ecosystem model in Fig.~\ref{fig:disc_ind2}. The exposure of compartment $1$ (PN-phyto) to the nitrogen input into the system at compartment $2$ (PN-hetero) during $[0,16]$, $e_{1_2}(0,t)$, is presented in Fig.~\ref{fig:real_expose} with the corresponding environmental nitrogen input, $z_2(t)$. The system efficiencies depicted in Fig.~\ref{fig:real_seff} have large fluctuations at about $t=7,8,9$, just as the impact of the Gaussian impulse in Case study~\ref{ex:hallam} (cf. Fig.~\ref{fig:hallam_ind_effs}). The unusual large fluctuations in the graphs of the composite \texttt{diact} system efficiencies should similarly indicate an excess amount of nitrogen input into the system. Indeed, the excess nitrogen inputs during this time period $[8,9]$ at compartments $4$ (DON) and $5$ (NOx), that is, $z_4(t)$ and $z_5(t)$, seem to be responsible for these fluctuations (see Fig.~\ref{fig:real_z}).

Biological activities increase in springs and summers and slow down during winters \cite{Christian1991}. Therefore, it is anticipated that the proportion of indirect flows varies seasonally, contributing more in the spring and less in the winter. There are attempts in the literature to demonstrate the seasonality in indirect effects that failed possibly due to the authors' indirect effect formulations as discussed in Section~\ref{sec:effects} \cite{Patten1985a,Patten1985,Borrett2006,Borrett2011,Ma2013}. The graphs of the composite indirect effect indices and the corresponding efficiencies presented in Fig.~\ref{fig:real_eff} and~\ref{fig:real_seff} can clearly capture the temporal system behavior as anticipated.

Interestingly, both flow- and storage-based cycling and acyclic effect indices, as well as the corresponding system efficiencies are oscillating in opposite phases and not well-ordered, as presented in Fig.~\ref{fig:real_eff} and~\ref{fig:real_seff}. As a matter of fact, the acyclic effect indices and the corresponding efficiencies oscillate in the opposite phase to all the other \texttt{diact} counterparts. Considering the supplementary nature of the acyclic and cycling flows, it is hypothesized that, possibly due to the slower biological activities during the winters, the transfer flows fall short of completing the nitrogen cycle and, therefore, the cycling flows decrease and the acyclic flows increase within the ecosystem during the winters \cite{Coskun2017SCSA}. These conclusions and interpretations imply that such precise quantitative analysis of ecosystems may lead to more ecological theoretical developments.

As demonstrated with the case studies in this section, the detailed information and inferences enabled by the system partitioning methodology cannot be obtained through the analysis of the original system by the state-of-the-art techniques.

\section{Discussion}
\label{sec:disc}

Nature is always on the move, and its systems are constantly changing to meet ever-renewing circumstances. Therefore, environment is not an easy concept to define and analyze mathematically. Although sound rationales have been offered in the literature for analysis of natural system dynamics under special cases, such as linear and static models, the need for dynamic analysis of nonlinear ecosystem models has always been present.

There have been a few attempts in recent decades to analyze dynamic ecological systems. Each of these attempts, however meaningful, has disadvantages as identified and comprehensively addressed by \cite{Coskun2017DCSAM}. The system decomposition theory and comprehensive methods proposed recently by \cite{Coskun2017NDP,Coskun2017DCSAM} potentially address the mismatch between the current static and computational methods and applied ecological needs. The system decomposition theory is based on the analytical and explicit, mutually exclusive and exhaustive system and subsystem partitioning methodologies. The dynamic system decomposition refines system analyses from the current static, linear, compartmental level to the dynamic, nonlinear, subcompartmental level. While the system partitioning determines the distribution of the initial stocks and environmental inputs, as well as the organization of the associated storages derived from the stocks and inputs within the system, the subsystem partitioning ascertains the distribution of arbitrary intercompartmental flows and the organization of the associated storages generated by these flows within the subsystems. The dynamic substorage and subthroughflow matrices, as well as the transient and \texttt{diact} flows and storages are formulated based on the system decomposition theory~\cite{Coskun2017DCSAM}. The subthroughflows and substorages determine the evolution of environmental inputs and associated storages individually and separately within the system. The transient and dynamic $\texttt{diact}$ transactions then determine the flows and storages transmitted along any given flow path and along all paths from one compartment, directly or indirectly, to another within the system. The system decomposition theory, therefore, decomposes the system to the utmost level.

Considering a hypothetical ecosystem modeling a food web with several interacting species for which the effects of a specific pollutant needs to be investigated, one of the most critical inquiries would be about the influence of the toxin in one species on any other in the system to address the potential harm. Assuming that the interspecific interactions are formulated deterministically, current mathematical methods can analyze only the direct effects of the pollutant through direct transactions. The dynamic effects of the toxin in one species indirectly through other species on another has never been formulated before. The system decomposition theory enables monitoring how an arbitrary amount of the pollutant travels along a chain of interactions, spread throughout the food web, transferred from one species directly or indirectly to another. The system decomposition, therefore, enables ascertaining the effects of the pollutant in one species, directly or indirectly, on any other in the ecosystem. The dynamic indirect effects measure is one of the multiple dynamic measures and indices introduced in the present manuscript.

The system decomposition theory constructs a base for the development of new dynamic system analysis tools. The time-dependent nature of these dynamic measures enables their time derivatives and integrals to be also formulated as novel ecosystem measures. Multiple dynamic measures and indices of matrix, vector, and scalar types are systematically formulated for the analysis of various attributes and characteristics of ecosystems in the present manuscript. More specifically, the flow- and storage-based, local-in-time and average, simple and composite \texttt{diact} effects, utilities, exposures, and residence times, as well as the corresponding system efficiencies, stress, resilience, and resistance are formulated systematically at both compartmental and subcompartmental levels. All of these mathematical system analysis tools are introduced analytically and explicitly as quantitative ecological indicators for the first time in literature. A mathematical technique for the quantitative classification and characterization of the main interspecific interaction types and the determination of their strength is also developed based on the \texttt{diact} effect and utility measures.

The \texttt{diact} effect measures and indices quantify the influence of system compartments directly or indirectly on other compartments, and the \texttt{diact} effect efficiencies and stress determine the efficiency of these influences. The \texttt{diact} utility measures and indices then quantify the relative influence of compartments on each other, and the \texttt{diact} utility efficiencies ascertain the efficiency of these relative influences. The \texttt{diact} exposures and residence times unravel the compartmental exposures to \texttt{diact} flows and compartmental activity levels, respectively. The time derivatives and integrals of these measures can detect disturbances and, therefore, dynamically quantify the system resilience and resistance.

The current measures and indices for ecological network analysis have significant shortcomings. For example, Finn's flow-based cycling index, FCI, developed four decades ago, has been an essential measure for ecosystem analysis but only for systems at steady state \cite{Finn1976,Finn1980}. The storage-based cycling effect index, SCI, recently introduced by \cite{Ma2014}, is also proposed for static systems. Various versions of static flow-based indirect effect indices have been formulated by multiple authors, but none of these seem to be precisely quantifying the indirect effects either, as discussed in Section~\ref{sec:effects} \cite{Leontief1966,Patten1985a,Patten1985,Ulanowicz1990b,Borrett2011,Ma2013}. Although derived with a different rationale, the static versions of the proposed flow- and storage-based, simple, compartmental, dynamic cycling indices are equivalent to the FCI and SCI, respectively \cite{Coskun2017SESM}. Unlike the analytical formulation of residence times in the proposed methodology, however, the residence times in the SCI definition are approximated through agent-based computational simulations. The proposed static indirect effect indices ascertain the corresponding phenomena more precisely than the previous static formulations in the literature \cite{Coskun2017SESM,Coskun2019ITR}. The system decomposition theory and comprehensive dynamic methods holistically addresses all the other shortcoming and disadvantages of the current static and computational techniques, measures, and indices. 

\section{Conclusions}
\label{sec:conc}

In the present manuscript, we systematically introduced multiple dynamic measures and indices of matrix, vector, and scalar types for the dynamic analysis of nonlinear compartmental systems in the context of ecology based on the system decomposition theory. These measures and indices for the \texttt{diact} effects, utilities, exposures, and residence times, as well as the corresponding system efficiencies, stress, resilience, and resistance are novel mathematical system analysis tools that serve as quantitative ecological indicators. A mathematical technique for the quantitative characterization and classification of main interspecific interaction types and the determination of their strength within food webs is also developed based on the \texttt{diact} effects and utilities.

The proposed dynamic system measures and indices extract detailed information about ecosystems' characteristics, functions, and behaviors. These measures and indices monitor the flow distribution and storage organization, quantify the \texttt{diact} effects and utilities of one compartment directly or indirectly on another, identify the system efficiencies and stress, measure the compartmental exposures to system flows, determine the residence times and compartmental activity levels, and ascertain the system resilience and resistance in the case of disturbances. Therefore, they may prove useful also for environmental assessment and management. Several case studies from ecosystem ecology are presented to demonstrate the efficiency and wide applicability of the proposed measures and indices.

The proposed dynamic methodology for the analysis of nonlinear ecosystems enhances the strength and extends the applicability of the state-of-the-art techniques and provides significant advancements in theory, methodology, and practicality. It serves, therefore, as a quantitative platform for testing empirical hypotheses, ecological inferences, and, potentially, theoretical developments. We consider that the proposed methodology brings a novel complex system theory to the service of urgent and challenging environmental problems of the day and has the potential to lead the way to a more formalistic ecological science.

\section*{Acknowledgments}
The author would like to thank Hasan Coskun for useful discussions and helpful comments that improved the manuscript.

\newpage
\bibliographystyle{siamplain}
\bibliography{ecology}

\end{document}